\documentclass[useAMS,usenatbib]{mn2e}
\usepackage{graphicx}
\usepackage{url}
\usepackage{amssymb}

\newcommand{\be}{\begin{equation}}
\newcommand{\ee}{\end{equation}}
\newcommand{\bea}{\begin{eqnarray}}
\newcommand{\eea}{\end{eqnarray}}

\newcommand{\nn}{\nonumber}

\newcommand{\dd}{\mathrm{d}}

\title[GRAVITY constraints on Sgr~A* flare models]{Distinguishing an ejected blob from alternative flare models at the Galactic centre with \textit{GRAVITY}}
\author[F. H. Vincent et al.]{F. H. Vincent,$^{1}$\thanks{E-mail: fvincent@camk.edu.pl} T. Paumard,$^{2}$ G. Perrin,$^{2}$ P. Varniere,$^{3}$ F. Casse,$^{3}$ F. Eisenhauer$^{4}$, 
\newauthor S. Gillessen$^{4}$ and P. J. Armitage$^{5,6}$ \\ 
$^{1}$Nicolaus Copernicus Astronomical Centre, ul. Bartycka 18, PL-00-716 Warszawa, Poland\\
$^{2}$LESIA-Observatoire de Paris, CNRS UMR 8109, UPMC, Universit\'e Paris-Diderot, 5 place Jules Janssen, 92195, Meudon, France\\
$^{3}$AstroParticule et Cosmologie (APC), Universit\'e Paris Diderot, 10 rue A. Domon et L. Duquet, 75205 Paris Cedex 13, France\\
$^{4}$ Max-Planck-Institut f\"ur Extraterrestrische Physik, 85748 Garching, Germany\\
$^{5}$ JILA, University of Colorado \& NIST, UCB 440, Boulder, CO 80309, USA\\
$^{6}$ Department of Astrophysical and Planetary Sciences, University of Colorado, Boulder, CO 80309, USA}
\begin{document}

\date{Accepted ... Received ... ; in original form ...}

\pagerange{\pageref{firstpage}--\pageref{lastpage}} \pubyear{2014}

\maketitle

\label{firstpage}

\begin{abstract}

The black hole at the Galactic centre exhibits regularly flares of radiation,
the origin of which is still not understood. In this article,
we study the ability of the near-future
\textit{GRAVITY} infrared instrument to constrain the nature of these
events. We develop realistic simulations of \textit{GRAVITY} astrometric data
sets for various flare models. We show that the instrument will
be able to distinguish an ejected blob from alternative flare models,
provided the blob inclination is $\gtrsim 45^{\circ}$, the flare brightest
magnitude is $14 \lesssim m_{\mathrm{K}} \lesssim 15$ and the flare
duration is $\gtrsim1$h$30$.
 
\end{abstract}

\begin{keywords}
Instrumentation: interferometers -- Astrometry -- Galaxy: centre -- Black hole physics
\end{keywords}

\section{Introduction}
\label{sec:intro}

It is very likely that the centre of our Galaxy harbours a supermassive
black hole, Sagittarius~A* (Sgr~A*), weighing $M=4.31\,10^{6}\,M_{\odot}$ at $8.33$~kpc from 
Earth~\citep{ghez08,gillessen09}.
A very interesting property of Sgr~A* is the emission of flares of radiation in the X-ray, 
near-infrared and sub-mm domains \citep[e.g.][]{baganoff01, ghez04, clenet05, yusefzadeh06b,eckart09,doddseden11}. 
The infrared events are characterized by an overall timescale of the order of one to two hours, 
and a putative quasi-periodicity of roughly 20 minutes; 
the luminosity of the source increases by a factor of typically 20 \citep[see][]{hamaus09}. 
The typical flux density at the maximum of the flare is of the order of 8 mJy \citep{genzel03, eckart09,doddseden11} 
which translates to approximately $m_{\mathrm{K}}=15$. 
The brightest infrared flare observed to date reached $m_{\mathrm{K}}=13.5$~\citep{doddseden10}.
The longest infrared flare observed lasted around $6$~h~\citep[or three times $2$~h if the light
curve is interpreted as many successive events,][]{eckart08}.

Various models are investigated to explain these flares: 
heating of electrons in a jet \citep[][hereafter \emph{jet model}]{markoff01}, 
a hotspot of gas orbiting on the 
innermost stable circular orbit (ISCO) of the black hole 
\citep[][hereafter \emph{hotspot model}]{genzel03,hamaus09},
the adiabatic expansion of an ejected synchrotron-emitting blob of plasma 
\citep[][hereafter \emph{plasmon model}]{yusefz06}, 
or the triggering of a Rossby wave instability (RWI) in the disc 
\citep[][hereafter \emph{RWI model}]{tagger06b,tagger06}.
Some authors question the fact that flares can be understood as specific events. 
The light curve fluctuation is then interpreted as a pure red noise, which
translates in a power-law temporal power spectrum with negative slope
\citep[][hereafter \emph{red-noise model}]{do09}. 
So far, no consensus has been reached and all models are still serious
candidates to account for the Galactic centre (GC) flares. 

The analysis of infrared Galactic centre flares will benefit in the very
near future from the astrometric data of the \textit{GRAVITY} instrument.
\textit{GRAVITY} is a second generation Very Large Telescope Interferometer 
instrument the main goal of which 
is to study strong gravitational field phenomena in the vicinity of 
Sgr~A*~\citep[][see also Table~\ref{grav} for a quick overview of the instrument's characteristics]{eisenhauer11}. 
 \textit{GRAVITY} will see its first light in May 2015.
As far as Galactic centre flares are concerned,
the exquisite astrometric precision of the instrument
is a major asset. \textit{GRAVITY} will reach a precision of the order of $10\, \mu$as, 
i.e. a fraction of the apparent angular size of the Galactic centre
black hole's silhouette~{\citep{bardeen73,falcke00}}, in only a few minutes of integration.
Such a precision will allow following the dynamics of the innermost Galactic centre, 
in the vicinity of the black hole's event horizon.
\begin{table}
 \centering
 \begin{minipage}{140mm}
  \caption{Main characteristics of the \textit{GRAVITY} instrument.}
  \label{grav}
  \begin{tabular}{@{}ll@{}}
 \hline
Maximum baseline length & 143~m\\ 
Number of telescopes & 4 (all UTs)\\
Aperture of each telescope & 8.2~m \\
Wavelengths used & 1.9 - 2.5~$\mu$m\\ 
Angular resolution &  4~mas \\
Size of the total\footnote{Containing the science target and a phase reference star} field of view& 2'' \\
Size of the science\footnote{Containing only the science target} field of view & 71~mas \\
\hline
\end{tabular}
\end{minipage}
\end{table}

The aim of this article is to \textit{determine to what extent can \textit{GRAVITY}
constrain the flare models from astrometric measurements only}.
We will thus only focus on the astrometric signatures of the various
models. Other signatures such as photometry or polarimetry~\citep[see e.g.][]{zama11} 
are not considered here and will be investigated in future papers.

Three broad classes of models can be distinguished depending on the 
astrometric signature of the radiation source:
\begin{itemize}
\item \emph{circular and confined} motion (hotspot model, RWI model);
\item \emph{complex multi-source} motion (red-noise model);
\item \emph{quasi-linear and larger-scale} motion (jet model, plasmon model).
\end{itemize}
The first class is characterized by a source in the black hole equatorial plane, following typically
a Keplerian orbit close to the black hole innermost stable circular orbit (ISCO). This source can
either be modelled by a phenomenological hotspot~\citep{hamaus09} or any
instability able to create some long-lived hotspot in the
disc, such as for example the RWI~\citep{tagger06b}. 
In the second class, many sources are located at different locations, ranging from the ISCO
radius $r_{\mathrm{ISCO}}$ to typically $10\,r_{\mathrm{ISCO}}$ (further out, the flux
becomes too small to be of significance for the flaring emission). 
These sources
follow non-geodesic trajectories. In the last class, the source is ejected along
a nearly linear trajectory out of the equatorial plane.
{We note that the two first classes may be called disk-glued models,
in the sense that they refer to physical phenomena inside the accretion structure
surrounding Sgr~A*. The third class is different as the source is ejected away from
the accretion structure.}

The point of this classification is that it is quite robust to changing the
details of the various models. This is the main reason supporting our choice of
considering astrometric data alone: even if the details of the physics of the various
models are not always fixed, the broad aspects of the source motion described in the
list above are quite firmly established. 

Our aim is to simulate \textit{GRAVITY} astrometric observations 
for these three classes of models and to determine whether 
these near-future data will
allow distinguishing between them. We will consider three models belonging
to the three classes above: the RWI, red-noise (hereafter RN)
and a blob ejection model.

Section~\ref{sec:flaremod} presents the three models and explains
how flare light curves are modelled. Section~\ref{sec:gravityobs} describes how
\textit{GRAVITY} astrometric data are simulated. Section~\ref{sec:distinguish} investigates
the ability of the instrument to distinguish between the three classes of models and Section~\ref{sec:conclu}
gives conclusions.

 
\section{Light curve and centroid modelling}
\label{sec:flaremod}

\subsection{General remarks}

We are interested in two observables associated with a GC flare: the light curve
and the centroid track of the source. The centroid is defined as the barycentre 
of the illuminated pixels on the observer's screen weighted by their specific intensity.
The main parameters of these observables are the flare duration and its brightest
magnitude in {\it K}-band, $m_{\mathrm{K}}$.
In the following we will simulate a 2~h-long infrared flare. This is a long flare, so
an optimistic choice, but still reasonable as such long flares have 
been observed e.g. by~\citep{eckart08,hamaus09}. The brightest magnitude will be
fixed to $m_{\mathrm{K}}=14$ or $m_{\mathrm{K}}=15$.

We describe a typical light curve of a GC infrared 
flare~\citep[see e.g. Figs.~6 and~8 of respectively][for a sample of observed light curves]{do09,hamaus09} as the superimposition
of two components: 

\begin{itemize}
\item an overall modulation, typically Gaussian, that increases the flux by
a factor of $2$ to $3$, with a total span of 2~h, i.e. around
4 ISCO periods for a Schwarzschild black hole of $4.31\,10^6 \, M_{\odot}$.
This Gaussian modulation is interpreted as the triggering, increase
and decrease of the model considered. In the following, we consider a factor
of $2$ Gaussian modulation;
\item a smaller-scale (periodic or not) fluctuating emission that can modulate the flux
at a level of a few to a few tens of percent, varying typically 
over one ISCO period.
\end{itemize}

{We note that flux factors of up to typically $20$ can be observed between the
quiescent level of Sgr~A* and the maximum of the brightest infrared flares~\citep{doddseden11}.
Here, the ground level of our simulations is defined as the faintest magnitude for which
GRAVITY has an astrometric error of the order of $10~\mu$as, thus $m_{\mathrm{K}}\approx15.5$.
From this level, a factor of $2$ to $3$ to reach the flare maximum is typical for bright infrared flares.
We also note that bright infrared flares, being observed during around $2$~h above $m_{\mathrm{K}}\approx15.5$,
have been reported few times in the recent years~\citep[e.g.][]{eckart08,doddseden11}.}

The RWI and RN models as described below can easily generate the small-scale fluctuating signal. 
As far as the RWI is concerned, as we are not modelling the triggering event of the 
instability~\citep[that can be modelled by the accretion of a blob of gas,][]{tagger06}, 
we do not have the
large-amplitude modulation in our simulations. 
In the RN case, the larger-amplitude modulation is assumed to be due to 
fluctuations
of the black hole accretion rate over a few ISCO periods. For both RWI and RN models, the larger-amplitude
modulation is reproduced by a Gaussian modulation.
On the other hand, the ejected blob model allows to generate
the overall modulation and needs no addition of a Gaussian signal. 
{As a consequence, we do not plan to compare the light curves
that we model to observed data, as the simple formulation we use for the RWI and RN
models does not allow yet to generate a light curve with a large enough flux increase.
We are here only interested in determining the source centroid wander for all models.
We believe that our centroid predictions are robust and can be compared to observations.
Indeed, the centroid tracks that we obtain are very similar to the ones
obtained by other authors, who used more sophisticated flare models. 
In particular, \citet{hamaus09} obtained
centroid tracks for a hotspot model that
can be compared with our RWI predictions. Our RN model predictions can be 
compared with Fig.~1 of~\citet{dexter12}: flux is distributed in a very similar
way for both simulations, which will translate to similar centroid tracks.
}

Let us note that, as far as \textit{GRAVITY} observations are concerned,
the Gaussian modulation is particularly important as a factor of $2$ in the flux
translates to a change of $0.75$ magnitude, which then translates to a
higher level of noise.

The following Sections describe the computation of the model-specific
emission and centroid track for the three classes of model under investigation.
Once the source emission and motion are known, maps of specific intensity
are computed using the open-source\footnote{Freely available at \url{gyoto.obspm.fr/}} 
\texttt{GYOTO} code~\citep{vincent11b}. Images of $300 \times 300$ pixels are computed
with a time resolution of $100~s$,
corresponding to the smallest integration time of \textit{GRAVITY}. Computing
the light curves and centroid tracks is then straightforward.

\subsection{Rossby wave instability}

The RWI can be seen as the form taken by the Kelvin-Helmholtz instability 
in differentially rotating discs and has a similar instability criterion.
For two-dimensional (vertically integrated) barotropic discs, the RWI can be 
triggered if an extremum exists in the inverse vortensity profile $\mathcal L$,
\begin{equation}
\mathcal L = \frac{\Sigma \Omega}{2 \kappa^{2}}\frac{p}{\Sigma^{\gamma}},
\label{eq:vortensity}
\end{equation}
where $p$ is the pressure, $\Sigma$ is the surface density, $\Omega$ is the rotation frequency, 
$\kappa^{2}$ is the squared epicyclic frequency, and $\gamma$ is the adiabatic index.
An extremum of vortensity can thus typically arise from an extremum of the epicyclic frequency, 
as is the case close to the ISCO of an accretion disc surrounding a black hole.
{Such an extremum is a pure geometrical effect, linked with our assumption
that the black hole is described by a pseudo-Newtonian potential mimicking
the Schwarzschild metric. Then, the epicyclic frequency profile shows an extremum
at some radius $r_{\mathrm{ext}}$ close to the ISCO. 
We assume that the accretion structure surrounding Sgr~A*
extends to small enough radii to reach $r_{\mathrm{ext}}$.}
Once it is triggered, the instability leads to large-scale spiral density waves and Rossby vortices.  
The dominant mode (i.e. the number of vortices) is dependent on the disc conditions~\citep{tagger06b}.

In this article, the RWI is simulated in the same way as in the two-dimensional simulations
of~\citet{vincent13} to which we refer for details of the numerical setup. Let us stress here
that these simulations are two-dimensional hydrodynamical simulations 
and use a pseudo-newtonian~\citet{paczynski80} potential. In this framework, the 
AMRVAC code~\citep{keppens11} is used to compute the surface density $\Sigma$ along the
disc as a function of the time coordinate $t$, as well as the emitting particles 4-velocity.
The time resolution of these hydrodynamical calculations is approximately equal
to $T_{\mathrm{ISCO}}/10$, where $T_{\mathrm{ISCO}}$ is the ISCO period. 
{This resolution was demonstrated to be sufficient for similar simulations
developed in~\citet{vincent13}.}
The total
computing time is of around $10 \,T_{\mathrm{ISCO}}$. Such a resolution allows to catch the
dynamical evolution of the instability. Going to higher resolution would only give access to details
of the dynamics that anyway are far beyond the reach of \textit{GRAVITY}.

{We highlight the fact that considering a geometrically thin disk, although
the accretion flow surrounding Sgr~A* is thick, is a reasonable assumption.
Indeed, we are interested here in simulating a single-hot-spot like structure, which
we assume is due to the RWI. We are not interested to investigate the detailed
signature of a RWI in Sgr~A* accretion flow. Thus, our first-order approximation
of a thin disk is sufficient for our purpose: determining the astrometric signature
of a hot-spot around Sgr~A*.}

The radiation mechanism is different from~\citet{vincent13}. Here we
follow the previous work of~\citet{falanga07} who developed simulations
of a magnetized disc subject to the RWI at the GC. As the density distributions
of the magnetized and unmagnetized RWI are similar, we use the same
expression for the frequency-integrated
intensity of infrared flare synchrotron radiation:

\be
I^{\mathrm{em}}_{\mathrm{IR}} \propto \Sigma^{5/2} \, r^{-3/4}
\ee
where $r$ is the coordinate radius labelling the accretion disc.

\begin{figure*}
\centering
	\includegraphics[width=5cm,height=5cm]{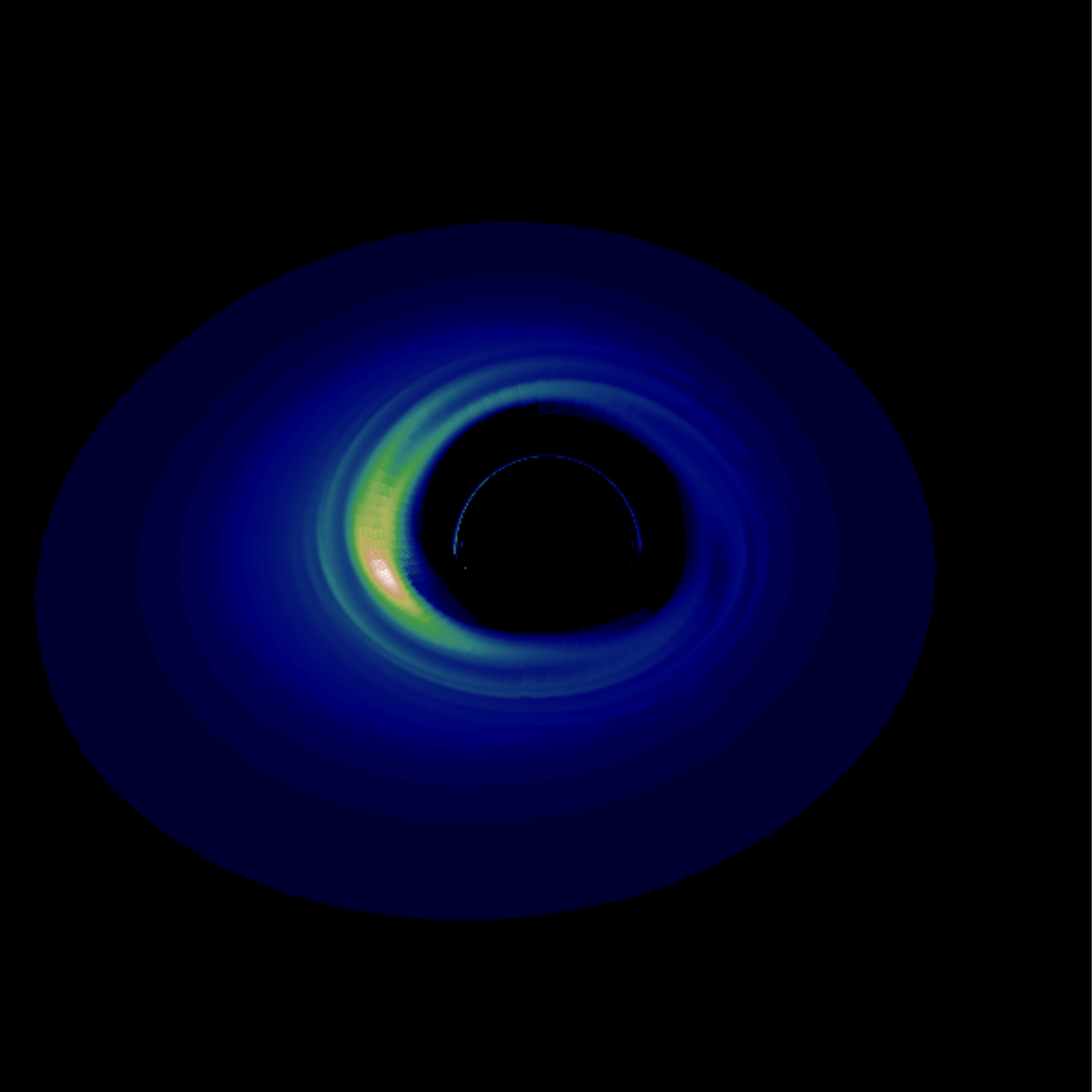}
	\includegraphics[width=5cm,height=5cm]{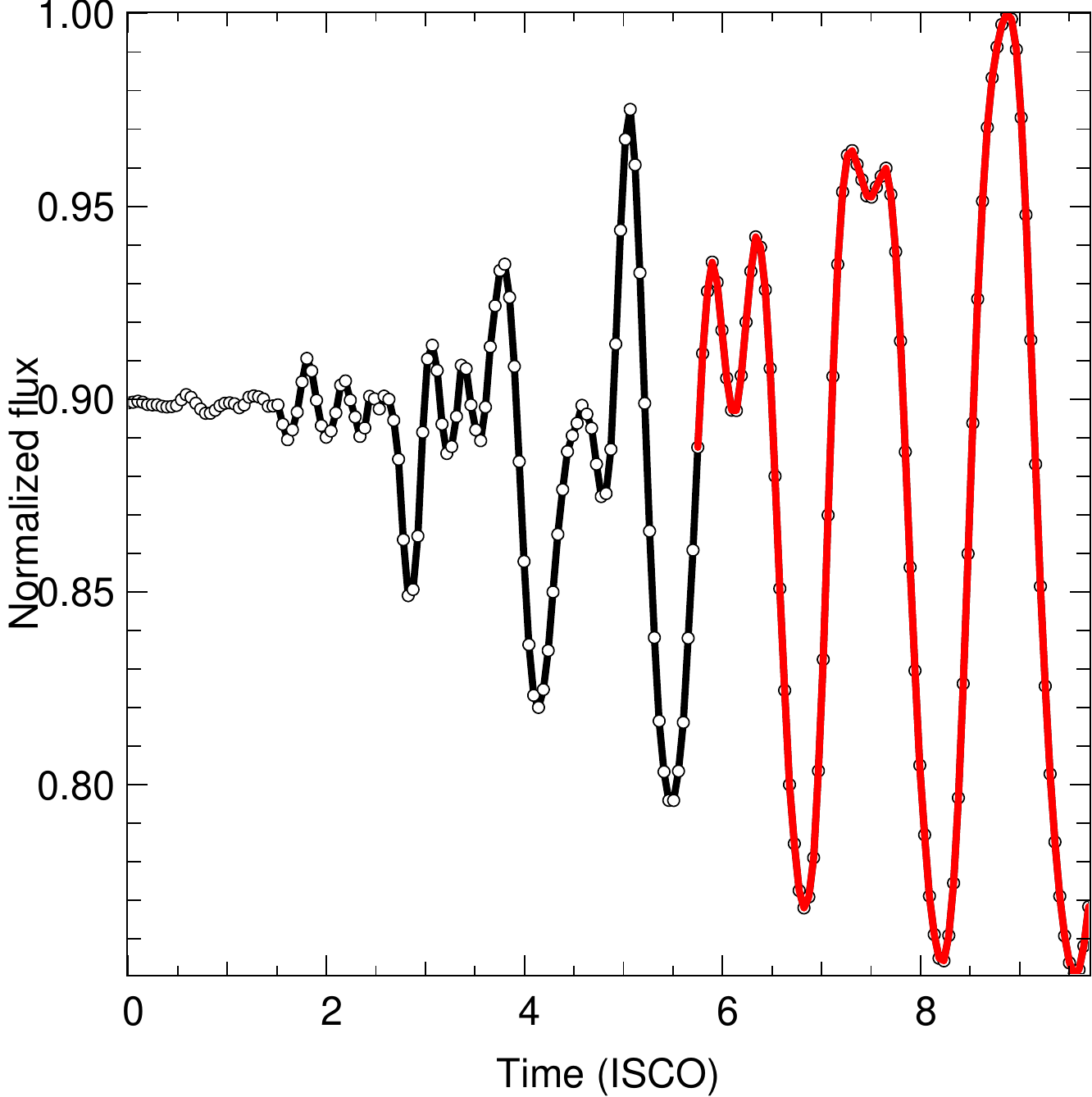}\\
	\includegraphics[width=5cm,height=5cm]{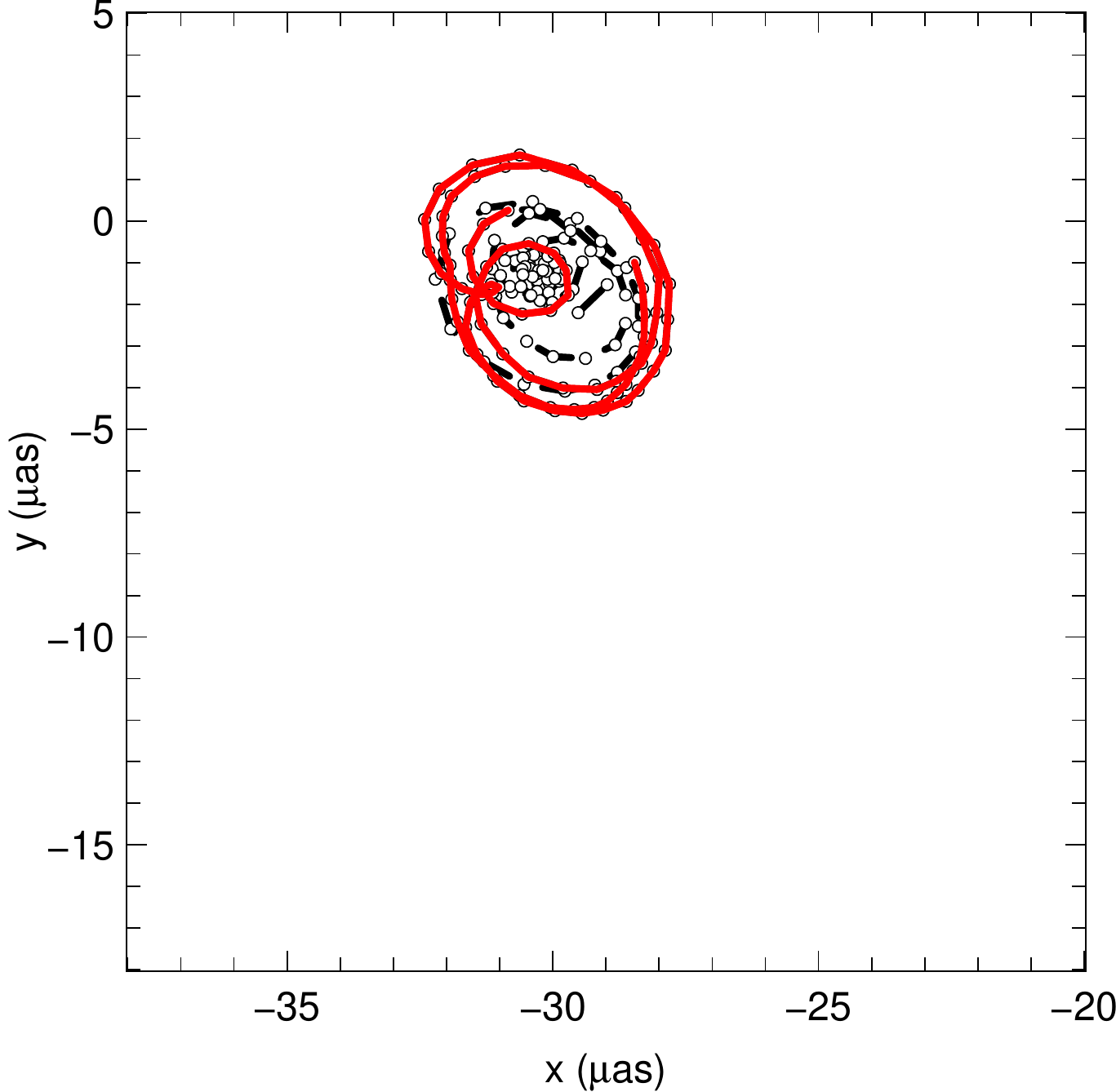}
	\includegraphics[width=5cm,height=5cm]{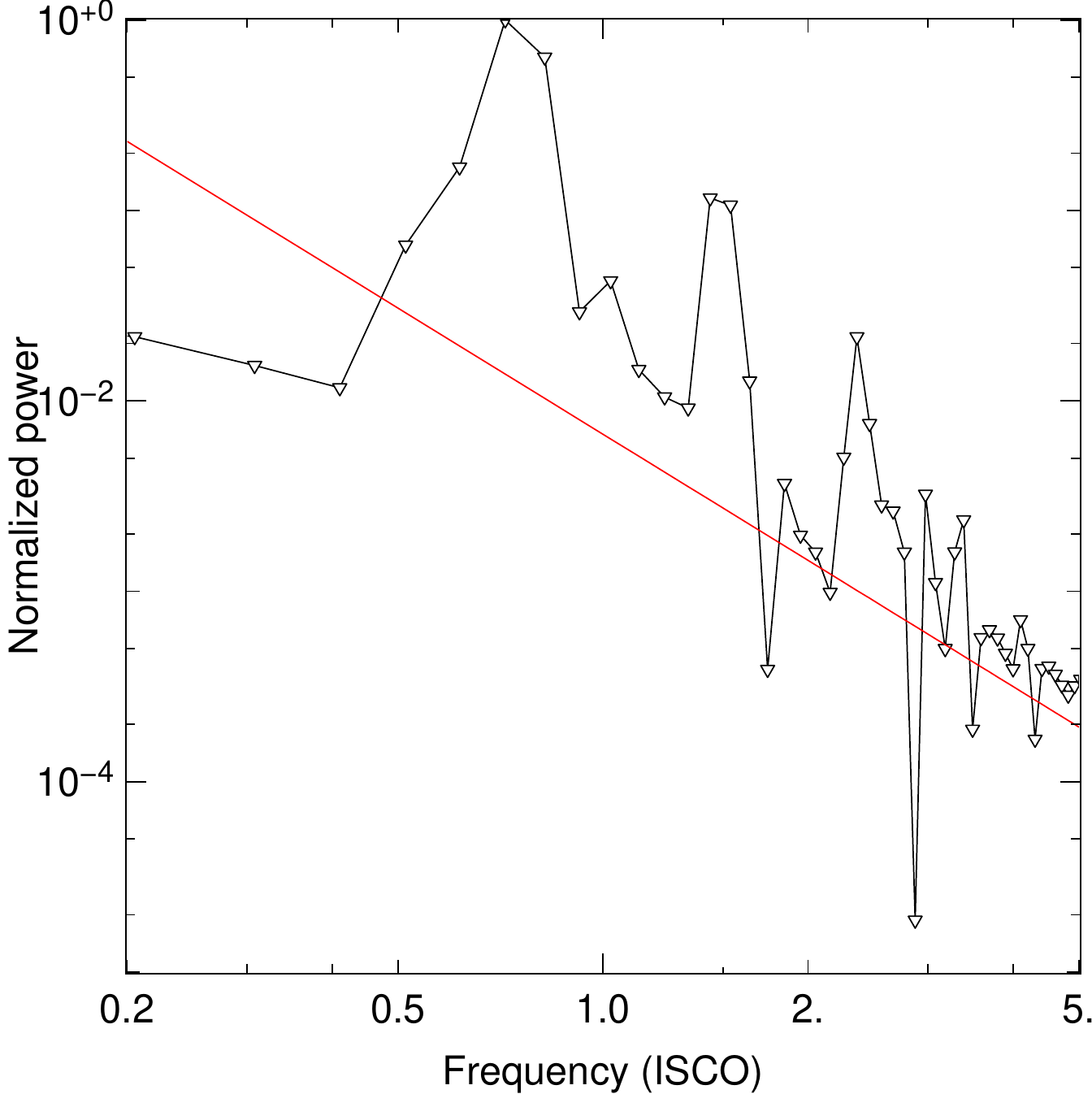}
	\caption{Image (upper left), light curve (upper right), centroid track (lower left) and power spectrum density (lower right) of a two-dimensional disc subject to the RWI,
		seen under an inclination of $45^{\circ}$. The thick red part of the light curve and centroid track shows 2~h of data, which is the duration
		we assume for the flare.
		The red line on the lower right panel shows a power-law spectrum with index $-2.2$, for
		comparison with Fig.~\ref{fig:fullRN}.}
	\label{fig:LCRWI}
\end{figure*}

Fig.~\ref{fig:LCRWI} shows the image of a two-dimensional disc with the RWI
fully developed, seen with an inclination of $45^{\circ}$, together with
the simulated light curve, centroid track and power spectrum density. 
These data last around $10 \,T_{\mathrm{ISCO}}$ which translates to around
$5$~h for a $4.31\,10^6 \, M_{\odot}$ Schwarzschild black hole. 
The details of the RWI modulation will depend on the flare's parameters.
Assuming that the
flare can be observed for around $2$~h only and that the small-scale modulation is strong, 
we select the last $2$~h of our
simulated data (in red color in Fig.~\ref{fig:LCRWI}), corresponding to the maximum modulation of the light curve. 
This choice is not very important as what matters most is the Gaussian modulation that
dictates the level of astrometric noise.
In the power spectrum density (right panel of Fig.~\ref{fig:LCRWI}), the modal signature of
the RWI appears clearly with peaks at once, twice and three times a fundamental
frequency which is a bit less than the ISCO frequency (as the instability is triggered
at a radius close but larger than the ISCO radius).

The red-color part of the light curve depicted in the upper right panel
of Fig.~\ref{fig:LCRWI} is then modulated by a Gaussian of full width
at half maximum adjusted in order to get a factor-two increase of the
light curve. The final light curve of the RWI flare model is illustrated in the
left panel of Fig.~\ref{fig:LCfinal}.

\subsection{Red noise}
\label{sec:RN}

Light curves of accretion discs subject to the magnetorotational instability (MRI)
leading to red-noise fluctuations were simulated by~\citet{armitage03}.
These authors performed magnetohydrodynamical (MHD) simulations of an accretion disc
surrounding a Schwarzschild black hole, described by a~\citet{paczynski80} pseudo-newtonian potential.
The vertically averaged magnetic stress is used as a proxy for the dissipation,
and light curves can be readily simulated from the MHD data.

Here we use the same MHD data and follow~\citet{armitage03} to express the
emitted intensity integrated over a range of infrared frequencies:
\be
\label{eq:emRN}
I^{\mathrm{em}}_{\mathrm{IR}} \propto \frac{B_r B_\varphi \left(r,\varphi,t\right)}{\langle B_r B_\varphi \rangle \left(r\right)} F_{\mathrm{PT}}\left(r\right)
\ee
where the $B_r B_\varphi$ terms are vertically integrated magnetic field components
and the average of the denominator is done over azimuthal direction $\varphi$ and time $t$. The 
$F_{\mathrm{PT}}(r)$ term is the standard expression of a two-dimensional disc's flux from~\citet{page74}.

We refer to the original work by~\citet{armitage03} for more details
on the numerical MHD simulations. Let us note that this work was developed for
relatively geometrically thin accretion disc. 
We use these results for the geometrically thick Sgr~A* accretion flow assuming
that the details of the vertical structure of the flow does not impact our results. As 
the astrometric data we model are only sensitive to the displacement of the centroid
projected on the observer's sky, we believe this is a sound assumption.
{Likewise, 
the heuristic emission law that we consider (Eq.~\ref{eq:emRN}) 
fits our goal of developing a first-order, simple red-noise model. We
do not claim that such a model is realistic enough to test the details
of a red-noise varying accretion structure surrounding Sgr~A*, but this
is not (yet) our goal.}
{We have checked that considering a different emission law than
the Page-Thorne prescription does not
change noticeably the centroid position.}

{We note that more realistic models of the accretion flow
surrounding Sgr~A* have been extensively developed in the past few 
years~\citep[e.g.][]{goldston05,noble07,chan09,mosci09,dexter10,shcher12}.
We will devote future work to developing our model in this way
in order to become able to study in detail the  
photometric, spectroscopic and polarimetric signatures 
of a red-noise accretion flow surrounding Sgr~A*.}

\begin{figure*}
\centering
	\includegraphics[width=5cm,height=5cm]{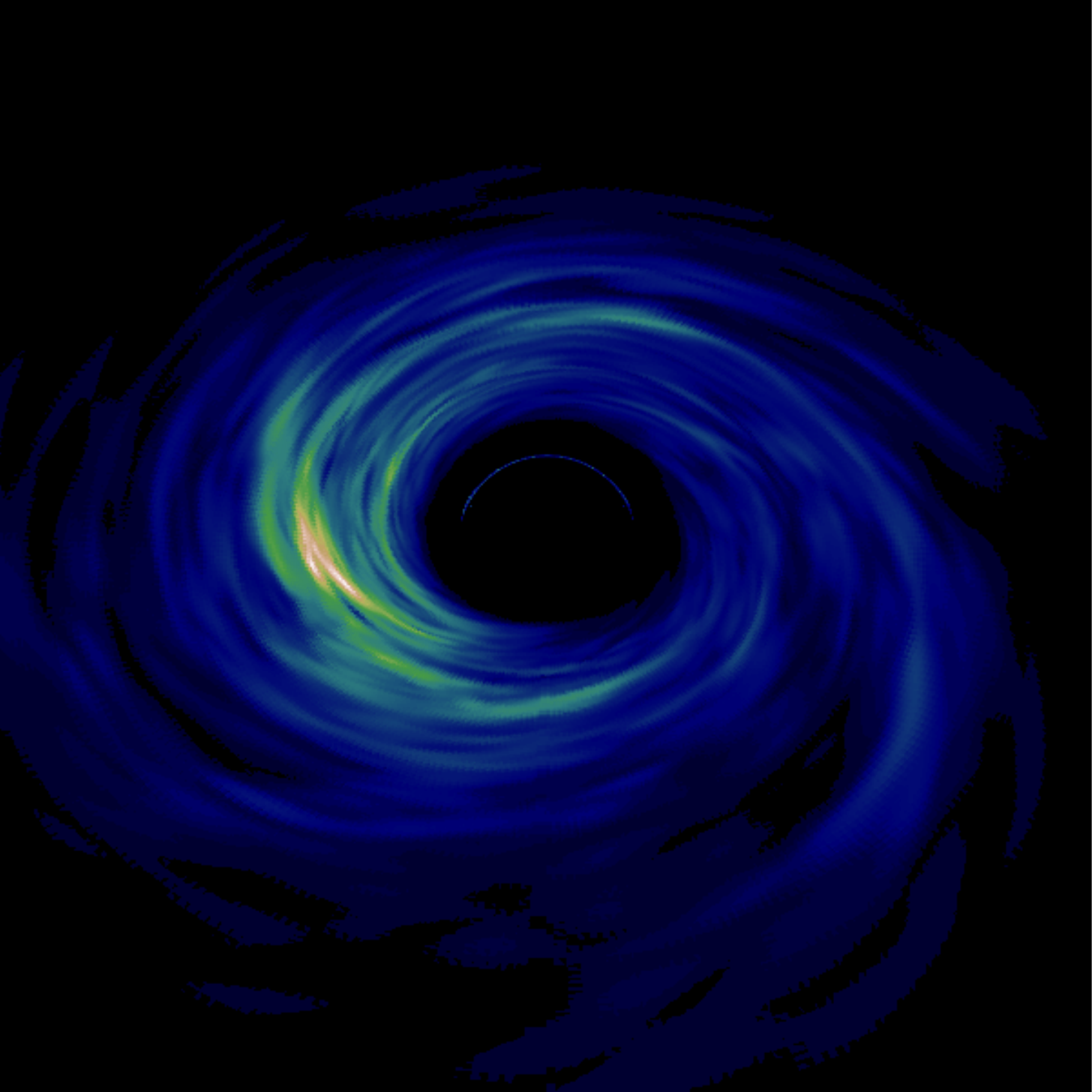}
	\includegraphics[width=5cm,height=5cm]{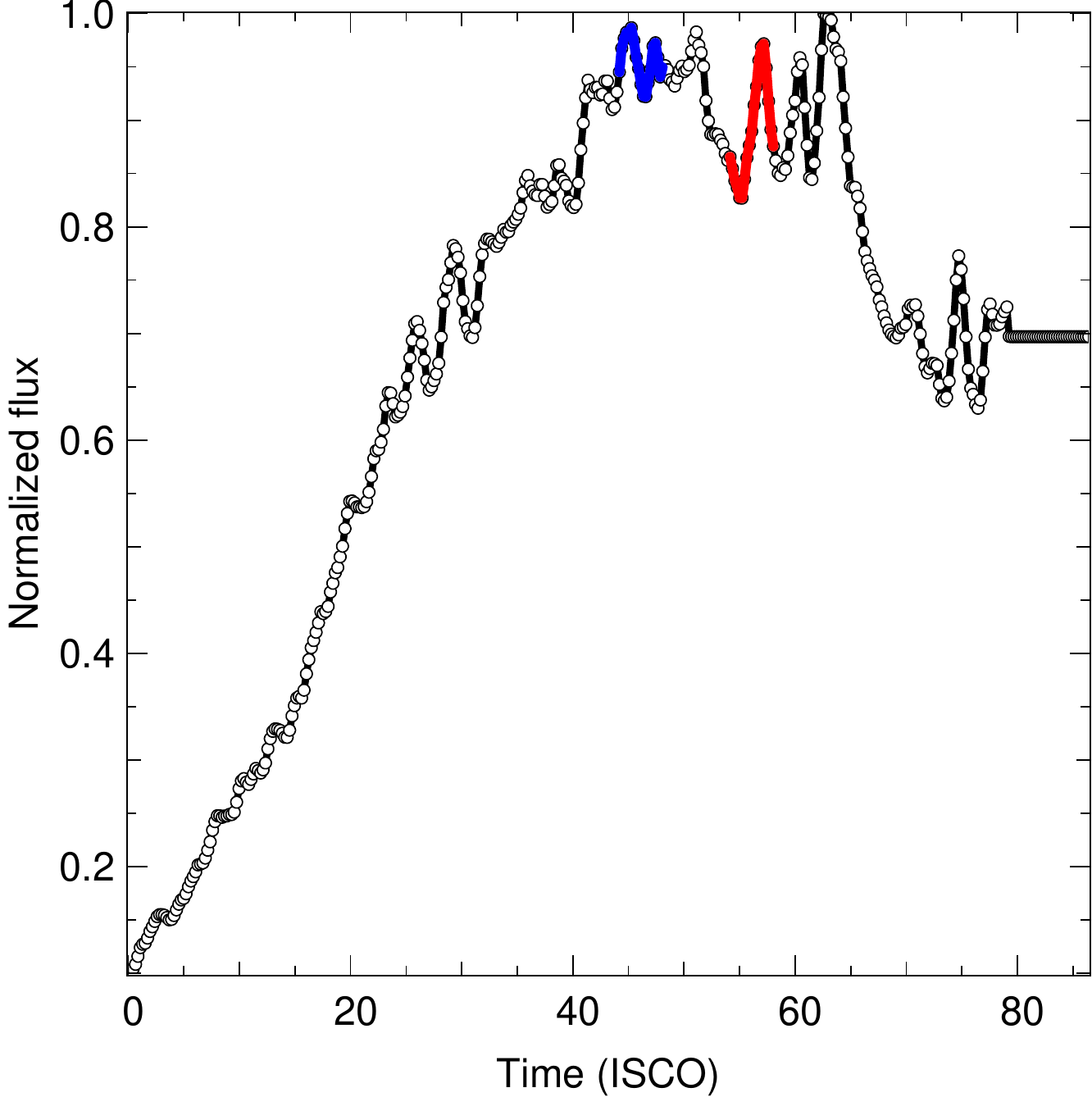}\\
	\includegraphics[width=5cm,height=5cm]{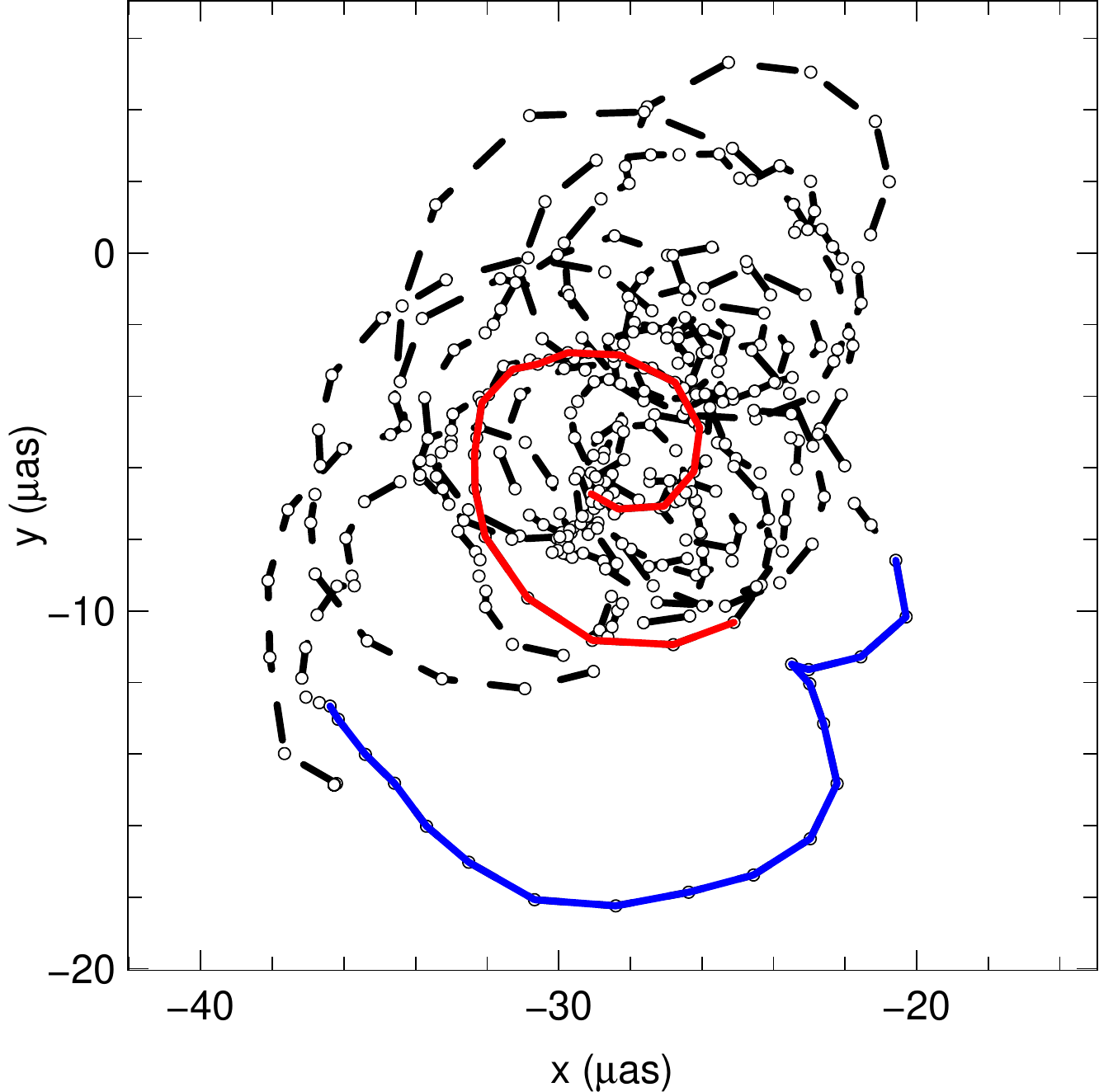}
		\includegraphics[width=5cm,height=5cm]{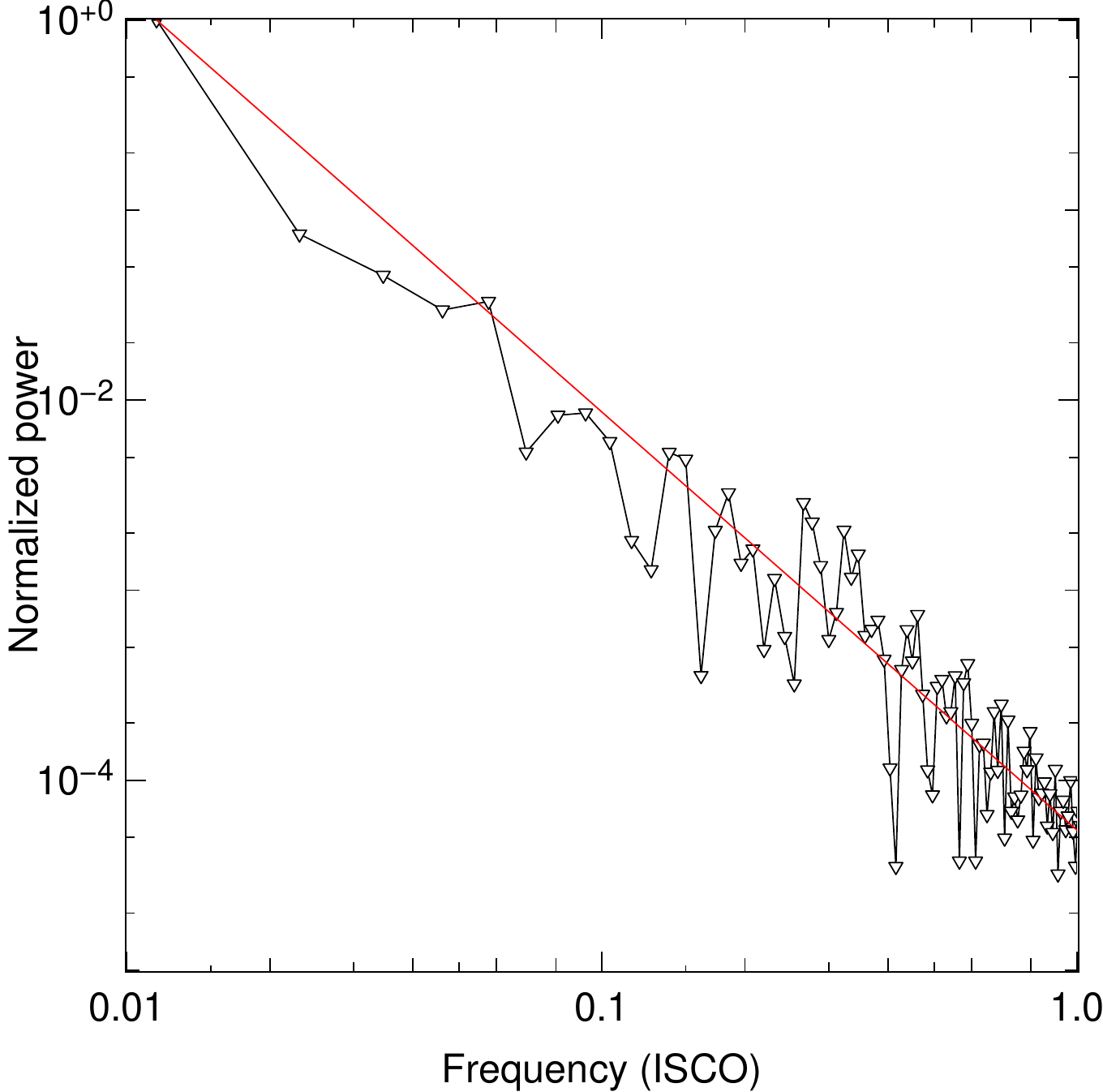}
	\caption{Image (upper left), light curve (upper right), centroid track (lower left) and power spectrum density (lower right) of a two-dimensional disc subject to red-noise fluctuations,
		seen under an inclination of $45^{\circ}$. Two $2$~h-long parts of the simulated data that are used for the GRAVITY
		simulations are shown in red and blue in the light curve and centroid track. The blue data correspond to a length of centroid track
		close to maximum
		while the red data correspond to a length close to minimum.
		The red line in the lower right panel shows a power law $f^{-2.2}$.
		}
	\label{fig:fullRN}
\end{figure*}

Fig.~\ref{fig:fullRN} shows an image of the red-noise model, the long-term light curve,
the associated centroid track and the power spectrum density for an inclination of $45^{\circ}$. The power
spectrum density shows a clear red-noise profile.
The increase (during $\approx 40\,T_{\mathrm{ISCO}}$) and subsequent decrease of the light curve
are due to variations of the accretion rate that are related to the simulation's initial conditions.
These long-term variations would not show in real data~\citep[see more details in][]{armitage03}.
As we are interested in only a few ISCO periods of data, we only consider the part of the light curve
where the long-term fluctuation of the accretion rate is least visible, i.e. between $\approx 40\,T_{\mathrm{ISCO}}$
and $\approx 60\,T_{\mathrm{ISCO}}$. In this range, we have selected two light curves of $2$~h,
corresponding to two extreme extensions of the centroid track of the source: one little extended
track (red in Fig.~\ref{fig:fullRN}) and one more extended track (blue). These two extreme cases show that
the extension of the centroid track can vary by a factor of $2$ depending on the initial conditions of the
red-noise model. In order to take this into account, $50\%$ of the simulated astrometric
data will use the little extended track case, and $50\%$ the more extended case.

The central panel of Fig.~\ref{fig:LCfinal} shows the Gaussian-modulated red-noise light
curve that will be used in the following. This light curve corresponds to the extended track
of the source (blue data in Fig.~\ref{fig:fullRN}).

\subsection{Ejected blob model}

We have developed MHD simulations of a blob of electrons ejected
along the $z$ axis (orthogonal to the equatorial plane) from an initial altitude of $z_{\mathrm{ej}}=6\,r_g$
along the $z$ axis, where $r_g=GM/c^2$. The MHD simulation is performed 
{using the AMRVAC code~\citep{keppens11}.}
The simulations are done in two dimensions
in the plane $(x,z)$ where $x$ is a coordinate in the equatorial plane. The
blob is then made three-dimensional by assuming it is axisymmetric.
The simulations are done using a pseudo-Newtonian potential. 
{The grid size is of around $0.1~r_g$.}
{The
magnetic field is assumed to remain constant, with a norm of $20$~G~\citep[see e.g.][]{liu04} and
vertical field
lines orthogonal to the equatorial plane. This is a very simplifying
assumption, but here as well as in the other models, we are not (yet)
interested in studying the details of an ejected blob. Our goal is to
determine the astrometric signature of a single blob of plasma (not
a continuous jet) when ejected linearly from the vicinity of Sgr~A*. The density
is computed as a function of time and}
the blob
is defined as the region where density is higher than $80\%$ of its
maximum value.
The typical radius of the initial blob is $R\approx 3\,r_g$. The main parameter of the model
is the ejection velocity that we choose equal to $v_{\mathrm{ej}}=0.4\,c$ which
is approximately equal to the Keplerian orbital velocity at ISCO.
For such an ejection velocity, the blob is not able to escape the black hole
attraction and ultimately falls back onto it (the escape speed for the blob we model, with
an initial altitude of $z_{\mathrm{ej}}=6\,r_g$,
is of around $0.45\,c$).
This choice of ejection velocity is dictated by the fact that the flare is assumed to last around 
$4$ ISCO periods. With our choice of ejection velocity, the complete simulation, from the
blob ejection to its coming back to the same location ($z=6\,r_{\mathrm{g}}$, the bottom of the simulation box), 
lasts around $5$ ISCO periods. For higher velocities, the blob escapes too fast to be seen for
$\approx 2~$h. Moreover, the observed centroid track will become longer
than the sub-millimeter constraint on Sgr~A* wandering~\citep{reid08}.
{As the plasmon model leads to sub-millimeter emission following infrared
emission, the infrared position wander should stay below the boundary of~\citet{reid08}.}
  For smaller velocities, the blob falls back onto the black hole too fast
as well. Thus our choice of velocity is well adapted for a long-duration flare.
The velocity is oriented along the $z$ axis and the blob is thus assumed
to be ejected orthogonally to the equatorial plane in the $z>0$ direction.
As the simulation box takes only into account the $z>6\,r_g$ region, the
interaction between the falling blob and the accretion disc is not modelled.
However, the rapidly decreasing blob's density will be much smaller 
than the disc's density at the impact, and
its temperature will as well be much smaller. It is thus unlikely that the
interaction of the falling blob with the disc would lead to any observable effect.

The blob is assumed to be emitting synchrotron infrared radiation.
Following the literature to compute this emission~\citep[see for instance][]{falcke95},
we express the emission and absorption coefficients according
to~\citep{rybicki79}

\bea
 j_\nu^{\mathrm{blob}} &=& C \,2 \pi^{3/2} \frac{\sqrt{3}q^3 B }{4\pi m c^2 (p+1)}\Gamma\left(\frac{p}{4}+\frac{19}{12}\right) \\ \nn
&&  \Gamma\left(\frac{p}{4}-\frac{1}{12}\right)\frac{\Gamma\left(\frac{7}{4}-\frac{p}{4}\right)}{\Gamma\left(\frac{9}{4}-\frac{p}{4}\right)}
 \left(\frac{2\pi m c \nu}{3 q B }\right)^{-(p-1)/2} \\ \nn
 \eea
and
 \bea
   \alpha_\nu^{\mathrm{blob}} &=& C \,2 \pi^{3/2} \frac{\sqrt{3}q^3}{8\pi m}\left(\frac{3q}{2\pi m^3 c^5}\right)^{p/2} B^{(p+2)/2} \\ \nn
 &&  \Gamma\left(\frac{3p+2}{12}\right)\Gamma\left(\frac{3p+22}{12}\right) 
  \frac{\Gamma\left(\frac{5+p}{4}\right)}{\Gamma\left(\frac{7+p}{4}\right)}
  \nu^{-(p+4)/2} \\ \nn
\eea
where $B$ is the magnetic field, $C$ and $p$ are defined by assuming that
the electrons energy density is a power law, $n(E)=C\,E^{-p}$, where $E$ is
the energy,  $q$ and $m$ are the electron's charge and mass. The gamma function $\Gamma$ is used. Here the 
pitch angle between the magnetic field and the electron velocity has been averaged,
as the electrons have a vertical bulk motion, but have random-oriented velocity
in the blob's frame. Let us note that this random velocity is much bigger than
the bulk velocity. Electrons have indeed Lorentz factors of at least $\gamma\approx100$
while the ejection velocity of $0.4\,c$ corresponds to $\gamma_{\mathrm{ej}}\approx1$.

Assuming equipartition between the magnetic and electrons energy densities,
\be
\frac{B^2}{8\pi} = C \int_{\gamma_m}^{\gamma_M} \gamma m c^2 \gamma^{-p} \dd \gamma
\ee
where $\gamma$ is the Lorentz factor and $\gamma_m$ and $\gamma_M$ are the
minimal and maximal Lorentz factors of the electrons. This immediately gives
\be
C = \frac{B^2}{8\pi m c^2} \frac{2-p}{\gamma_M^{2-p}-\gamma_m^{2-p}}
\ee
for $p\ne2$. In the following, we assume $p=4$, $\gamma_m=100$
and $\gamma_M=100\,\gamma_m$.

The radiative transfer equation can thus be straightforwardly 
integrated in the blob, which is done by the \texttt{GYOTO} code.
As only the bulk velocity of the blob is known, and not the instantaneous
random velocity of the emitting electrons, it is not possible to compute the
redshift factor $g$ related to the emitter's 4-velocity, such that
the observed and emitted specific intensities are related by
$I_{\nu}^{\mathrm{obs}} = g^4 I_{\nu}^{\mathrm{em}}$. This factor is thus
imposed to be constant for all \texttt{GYOTO} screen pixels. This boils
down to assuming that the emitter is on averaged at rest in the blob's frame,
which is true as in this frame the emitter's velocity has random orientation.

\begin{figure*}
\centering
	\includegraphics[width=5cm,height=5cm]{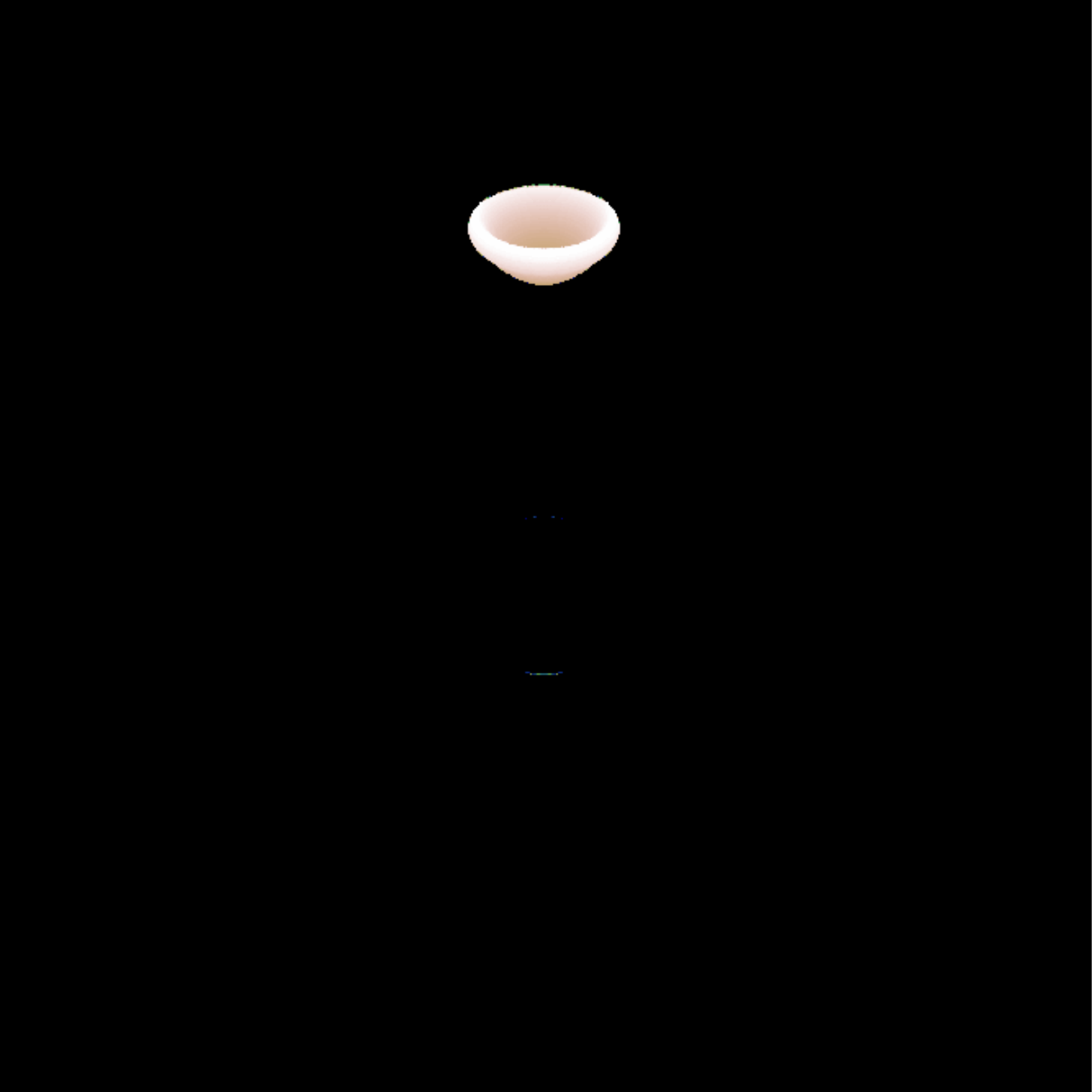}
	\includegraphics[width=5cm,height=5cm]{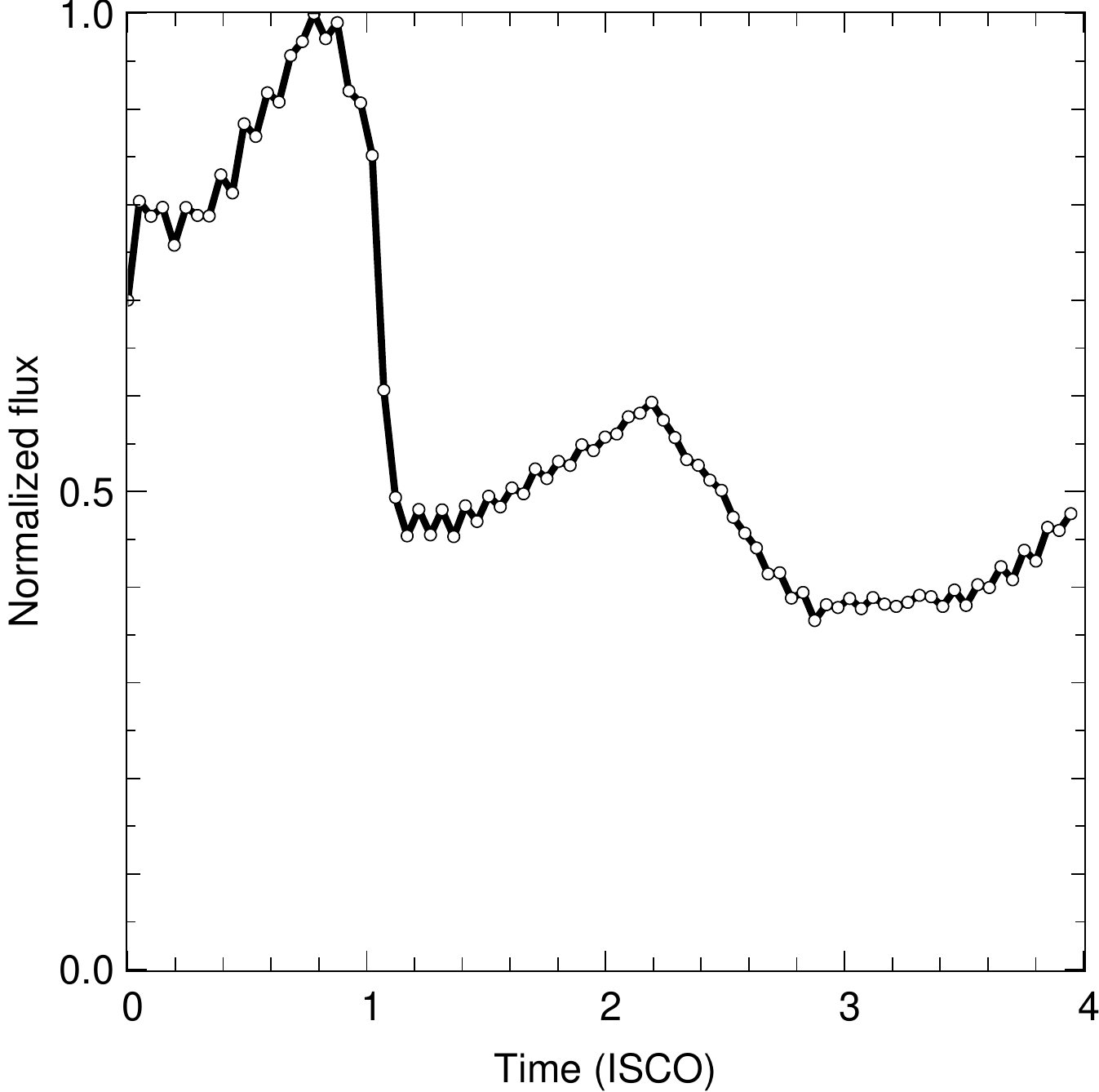}
	\includegraphics[width=5cm,height=5cm]{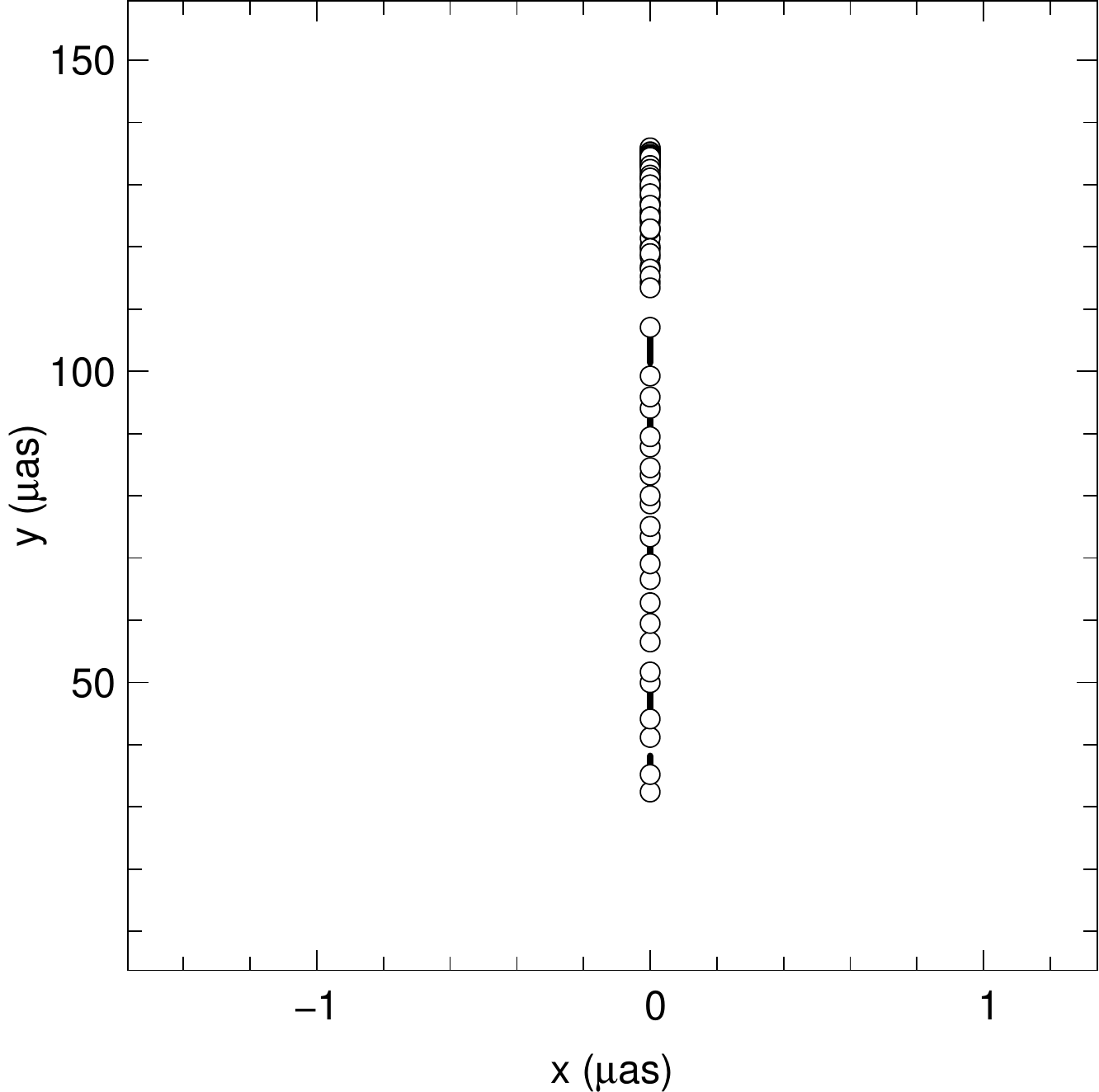}
	\caption{Image (left), light curve (centre) and centroid track (right) of the ejected blob,
		seen under an inclination of $45^{\circ}$. In the left panel, the blob is close to its
		maximal altitude, before falling down onto the black hole. Its typical radius
		at this moment is of around $4~r_g$.
		}
	\label{fig:fullBlob}
\end{figure*}

Fig.~\ref{fig:fullBlob} shows a ray-traced image, the light curve and the
centroid path of the blob model as seen from an inclination of $45^{\circ}$.
The variation of the light curve is mainly due to the change of projected
area of the blob. It first expands, reaches a maximum, and is then 
divided into two sub-blobs due to tidal effects, that fall back onto the black hole.
Let us note that the total angular displacement of the blob on the observer's
sky is of around $100~\mu$as in approximately $2$~h. Such a displacement 
is of the same order as the detected level of Sgr~A* wandering by sub-millimeter 
interferometry~\citep{reid08}. Higher ejection velocities would lead to exceeding this limit.

\vspace{1cm}

\begin{figure*}
\centering
	\includegraphics[width=5cm,height=5cm]{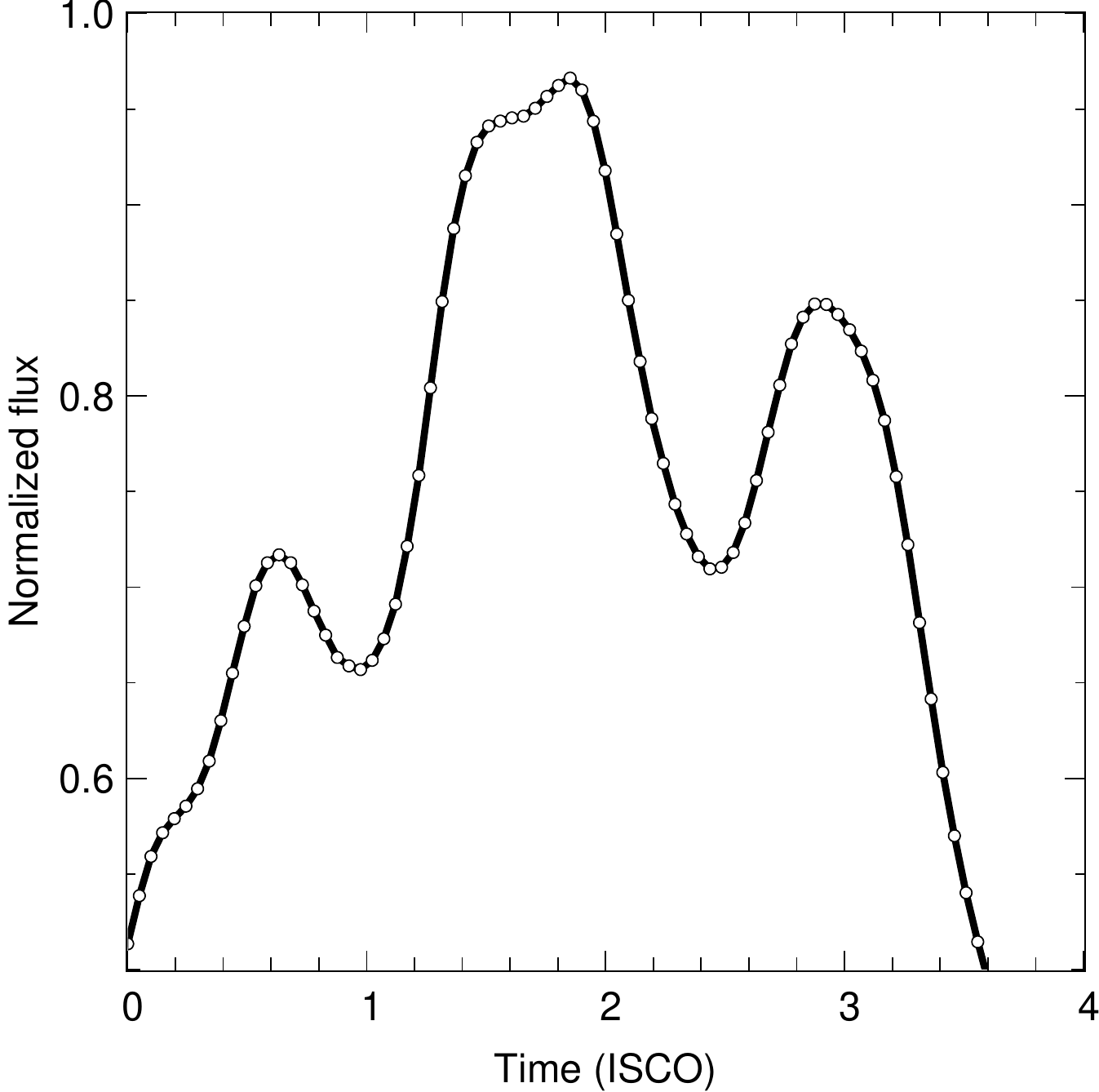}
	\includegraphics[width=5cm,height=5cm]{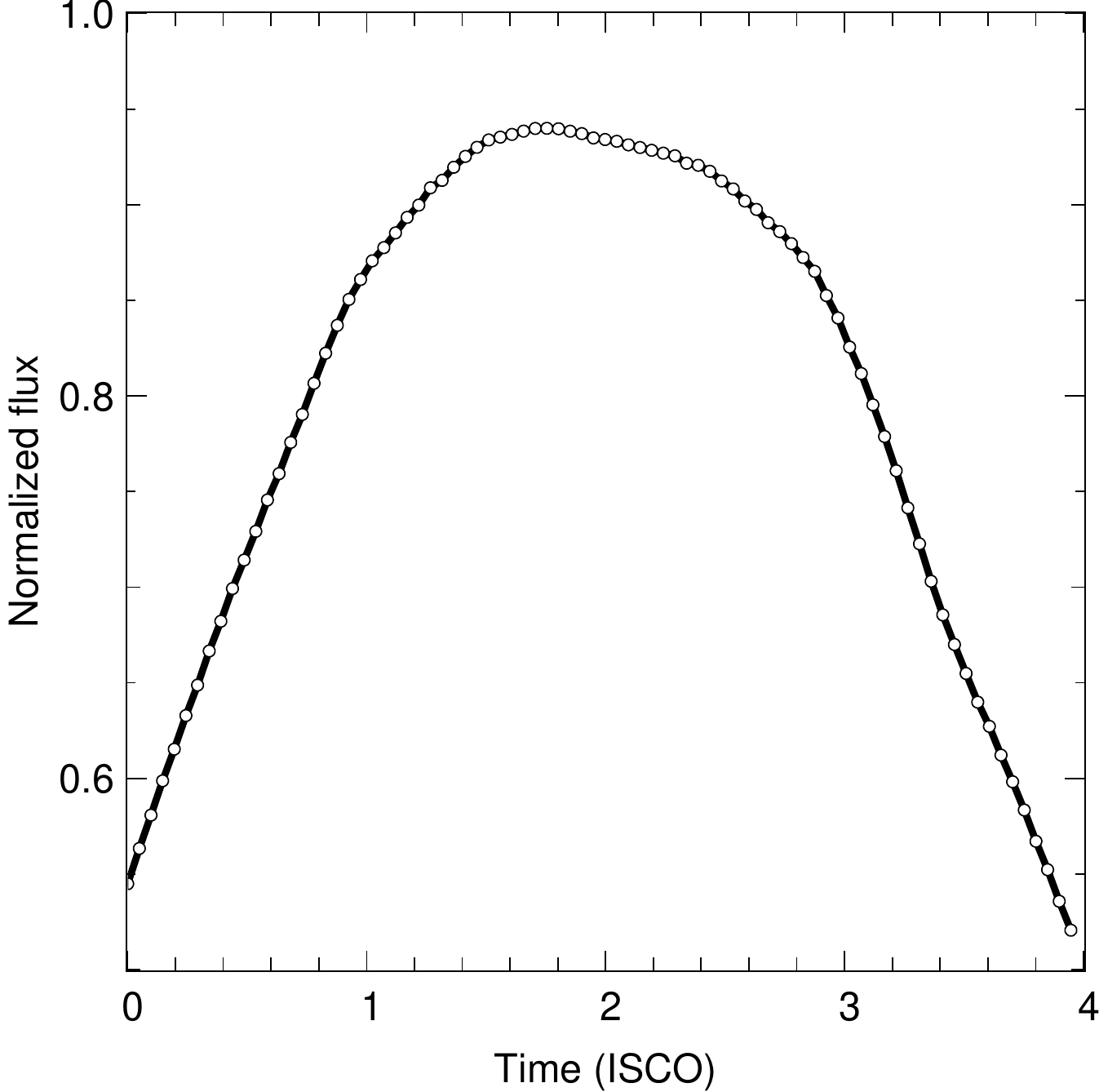}
	\includegraphics[width=5cm,height=5cm]{LC_Blob.pdf}
	\caption{Light curves of the three models investigated, as used for the \textit{GRAVITY} simulations: RWI (left), red noise (centre), blob (right). 
		The RWI and RN light curves from Figs.~\ref{fig:LCRWI} and~\ref{fig:fullRN} are here modulated by a Gaussian. 
		The time is given in units of the ISCO period, which is $\approx 30$~min for a Schwarzschild black hole of 
		$4.31\,10^6 \, M_{\odot}$. These three light curves look quite different, but observed infrared data show a large
		variety~\citep[see e.g.][]{do09,hamaus09}.}
	\label{fig:LCfinal}
\end{figure*}

Fig.~\ref{fig:LCfinal} shows the three light curves for all flare models, at $45^{\circ}$ of inclination,
that will be used to generate \textit{GRAVITY} simulations. 
{The light curves look different from one model to the other: it may seem likely that models could
be distinguished only by studying the observed flux. However, as stated above, our RWI and RN models
are restricted to model the small-scale variation of the flux and are not meant to reproduce all aspects
of observed light curves. As a consequence, our simulated light curves should not be compared
to observed data. Their \textit{only} goal in this article is to give a realistic evolution of the K-band magnitude
of the source, and thus of the \textit{GRAVITY} astrometric error.}

Next Section describes the \textit{GRAVITY} data acquisition.

\section{Simulation of a \textit{GRAVITY} observation}
\label{sec:gravityobs}

\subsection{Simulating \textit{GRAVITY} astrometric data}

The simulation of \textit{GRAVITY} astrometric data given some
model of emission in the vicinity of Sgr~A* is described in detail
in~\citet{vincent11}. We will only briefly summarize the main steps 
here for convenience.

The light curve and centroid track of some model being known,
a set of angular positions on sky and intensities $(x_{\mathrm{th}}(t),y_{\mathrm{th}}(t),I_{\mathrm{th}}(t))$ 
are available for all observing times $t$.
Here the subscript indicates that these quantities are theoretical,
computed by our set of numerical codes described above.
The $y$ coordinate is along the normal to the equatorial plane,
while $x$ is orthogonal to $y$.

From this, a set of theoretical complex visibilities
\begin{equation}
V_{\mathrm{th}}\left(t,u,v\right) =  I_{\mathrm{th}}(t)\,e^{-2\,i\,\pi\,\left(u \,x_{\mathrm{th}}(t)+v\,y_{\mathrm{th}}(t)\right)}
\end{equation}
can be computed for all spatial frequencies $(u,v)$. Given that
\textit{GRAVITY} uses four telescopes (i.e. six baselines) 
and five spectral channels in its most sensitive mode,
there are $30$ pairs of $(u,v)$ points and hence $30$ values of theoretical
complex visibilities associated to one given set 
$(x_{\mathrm{th}}(t),y_{\mathrm{th}}(t),I_{\mathrm{th}}(t))$.

Realistic noise is added to these theoretical 
visibilities~\citep[see the detailed description of the sources of noise in][and the short presentation 
of detection noise in the Appendix]{vincent11}
in order to get the simulated observed visibilities $V_{\mathrm{obs}}\left(t,u,v\right)$.

The retrieved astrometric data $(x_{\mathrm{obs}}(t),y_{\mathrm{obs}}(t))$
are then found by fitting the observed visibilities, assuming intensity can be
measured with $10\%$ accuracy, {which is the typical photometric accuracy of GRAVITY}. 
The fitting procedure is done using the Levenberg-Marquardt
algorithm as implemented in the \texttt{Yorick} language by the \texttt{lmfit} routine.
The astrometric error on this retrieved position is found as a function of the source magnitude 
at time $t$ by using the \textit{GRAVITY} astrometric error study presented in~\citet{vincent11}.
As the orientation of the normal to the black hole's equatorial plane with respect to the axes of
the instrument's point spread function (PSF) is not known, we assume an equal astrometric
error in the two orthogonal direction $x$ and $y$ (that is to say, we assume that the
normal to the equatorial plane makes an angle of $45^{\circ}$ with respect to the PSF axes).

\subsection{Simulating a flare observation by \textit{GRAVITY}}

We consider one night of observation chosen on 2014 April 15. 
Choosing the $15^{\mathrm{th}}$ day of the month is purely arbitrary. 
However, the choice of the month has an impact on the total time during which the GC is observable. 
On 2014 April 15, the GC is observable for a little more than 5 hours (as a comparison, it would be observable for 9 hours in June). 
As the duration of a flare is assumed to be $2$~h, it is possible to observe the whole event.

The integration time of the instrument is set to $100$~s, allowing to follow the dynamics of the flare. 
The Earth rotation during one elementary block is supposed to be negligible because the trace 
in the $(u,v)$ plane during one integration block is perfectly linear and can be replaced by its average. 
 
The light curve is scaled by the brightest magnitude of the flare, 
which is chosen to be $m_{\mathrm{K}}=14$ or $m_{\mathrm{K}}=15$. 
The brightest magnitude is still a realistic choice for a brigth flare, as the most powerfull such event 
has been observed at $m_{\mathrm{K}}=13.5$~\citep{doddseden11}.
\begin{figure*}
\centering
	\includegraphics[width=5cm,height=5cm]{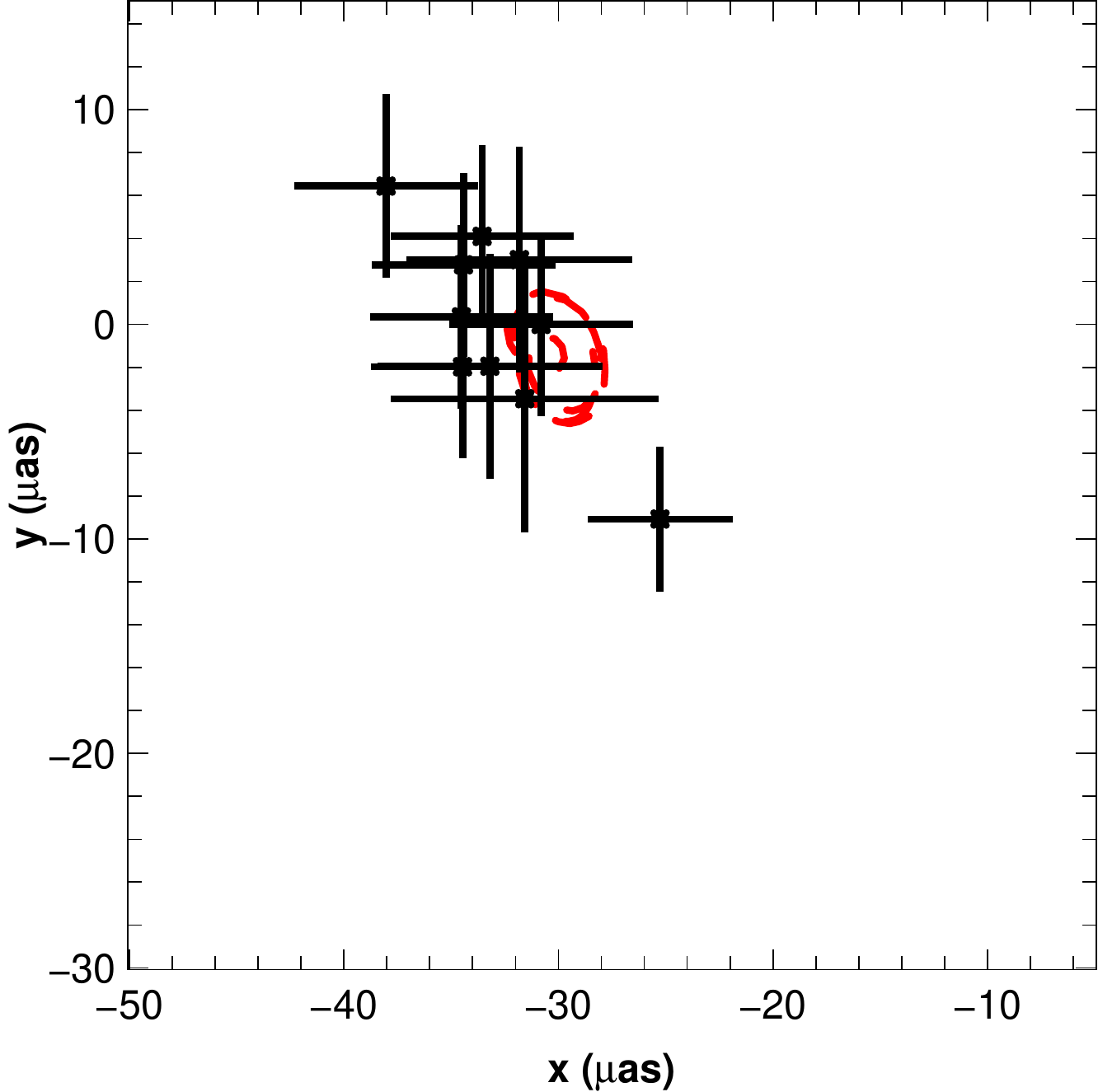}
	\includegraphics[width=5cm,height=5cm]{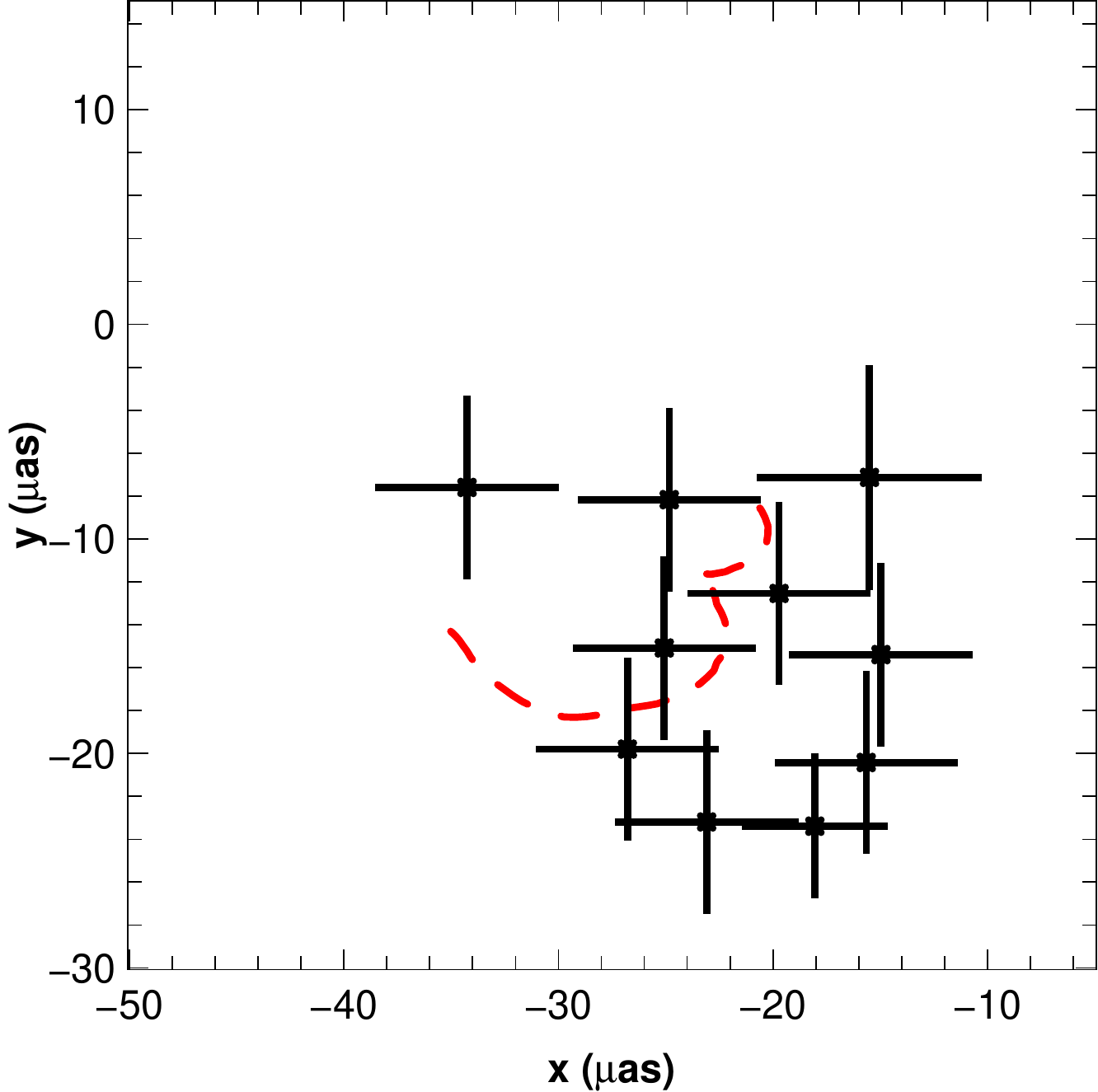}
	\includegraphics[width=5cm,height=5cm]{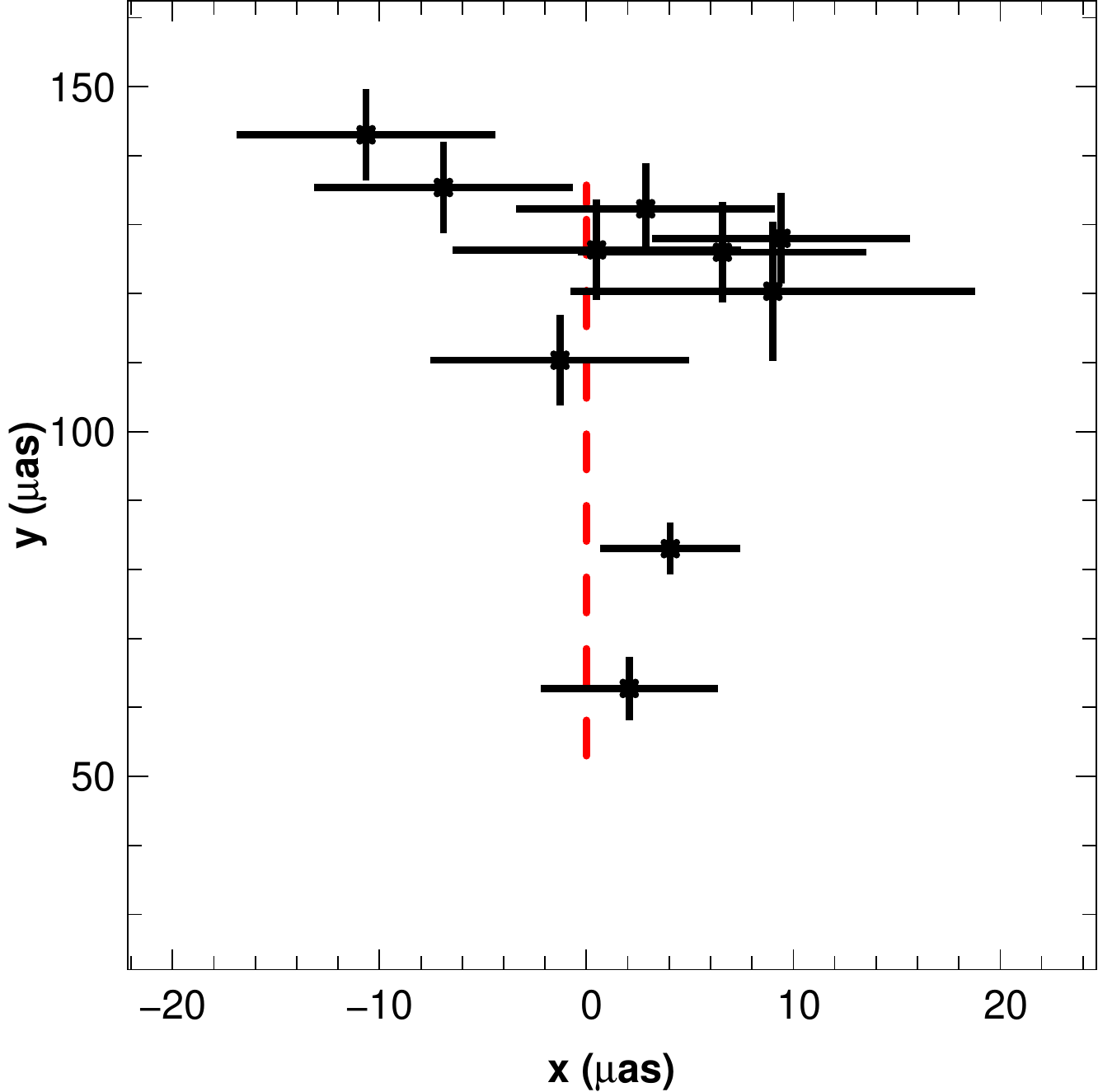}
	\caption{\textit{GRAVITY} observation simulation of one night with a flare due to the RWI model (left),
		the red-noise model (centre) or the blob model (right), with inclination $45^{\circ}$. The dashed red line is the theoretical centroid track. The error
		bars take into account the realistic astrometric performances of the instrument. The axes are graduated
		in microarcseconds. Mind the different vertical scaling of the panels.}
	\label{fig:flare}
\end{figure*}

Fig.~\ref{fig:flare} shows a simulation of one such night of observation by \textit{GRAVITY}
for all flare models considered here, the inclination being
assumed to be $45^{\circ}$. 
The size of the error bars are not all the same as they depend on the magnitude of the flare
at the time of observation. 
If the brightest magnitude is $m_{\mathrm{K}}=14$, it will decrease to reach at the lowest level
$m_{\mathrm{K}}=14.75$, which translates to a precision for a single observation (no averaging)
of respectively $8~\mu$as and $16~\mu$as~\citep{vincent11}.
The retrieved data are
averaged in order to get a better precision and only $10$ astrometric data points are conserved
from the initial $\approx 70$ retrieved data points, hence a factor $\approx 2.5$ better in precision.
This is the reason why the error bars are better than the single-observation precision.


\section{Distinguishing an ejected blob from \textit{GRAVITY} astrometric data}
\label{sec:distinguish}

\subsection{Directional dispersion}

In this Section we investigate whether the different flare models can
be distinguished by computing the dispersion of the \textit{GRAVITY} retrieved positions.
The directional dispersions $\sigma_x$ and $\sigma_y$ are defined as:

\bea
\label{eq:disp}
\sigma_x &=& \sqrt{\frac{1}{N-1}\sum \left(x-x_{\mathrm{moy}}\right)^2}, \\ \nonumber
\sigma_y &=& \sqrt{\frac{1}{N-1}\sum \left(y-y_{\mathrm{moy}}\right)^2}, \\ \nonumber
\eea
where $N$ is the number of averaged data, $x_{\mathrm{moy}}$ and $y_{\mathrm{moy}}$
are the averaged of the $x$ and $y$ retrieved coordinates.
One value of directional dispersion $(\sigma_x$, $\sigma_y)$ is computed for every simulated night of observation.
In the framework of a Monte Carlo analysis, we simulate $1000$ nights of observation
for all three models. 
The histograms of the resulting dispersions are analysed in the following Section.

\subsection{Results}

\begin{figure*}
\centering
	\includegraphics[width=5cm,height=5cm]{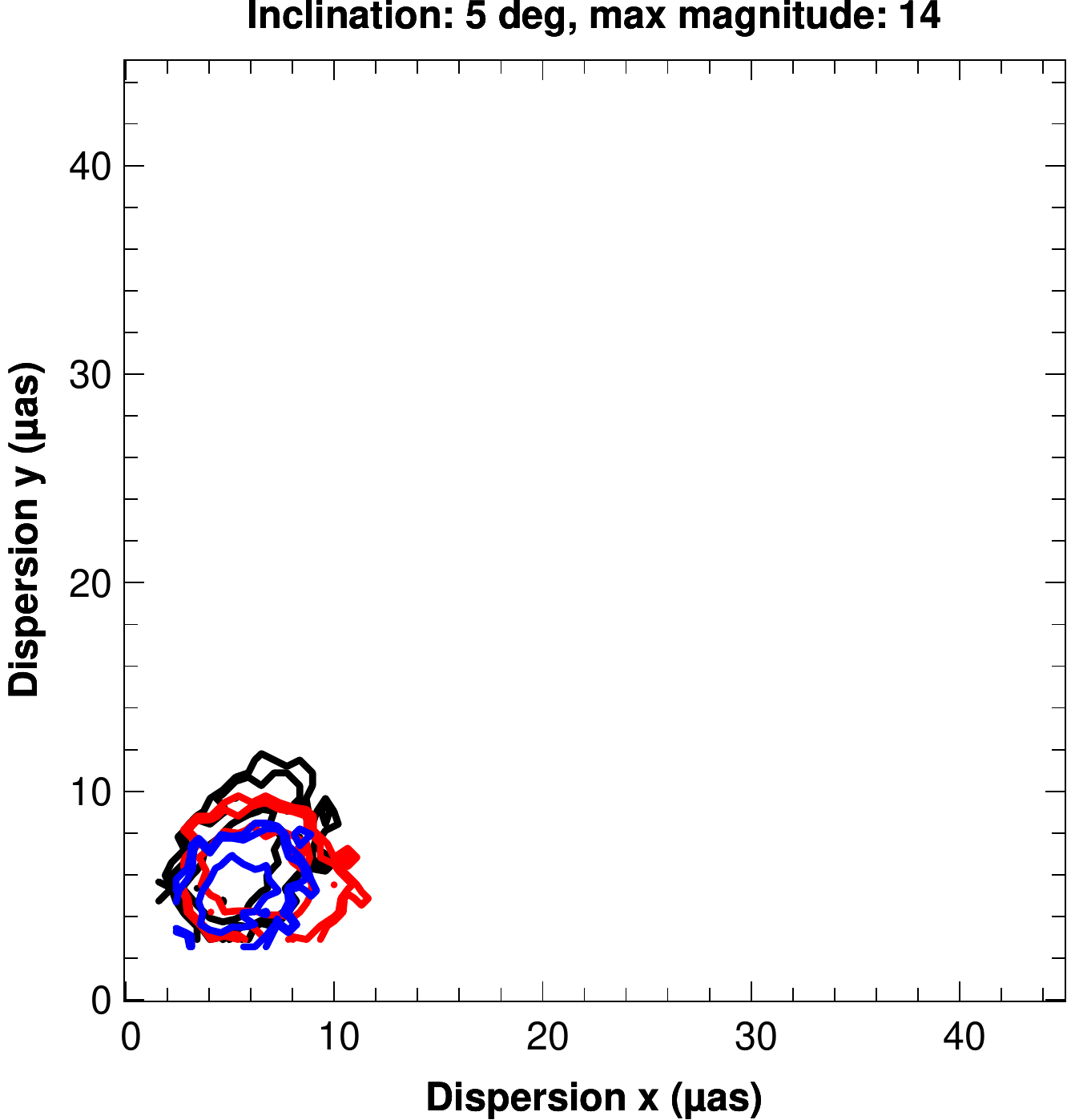}
	\includegraphics[width=5cm,height=5cm]{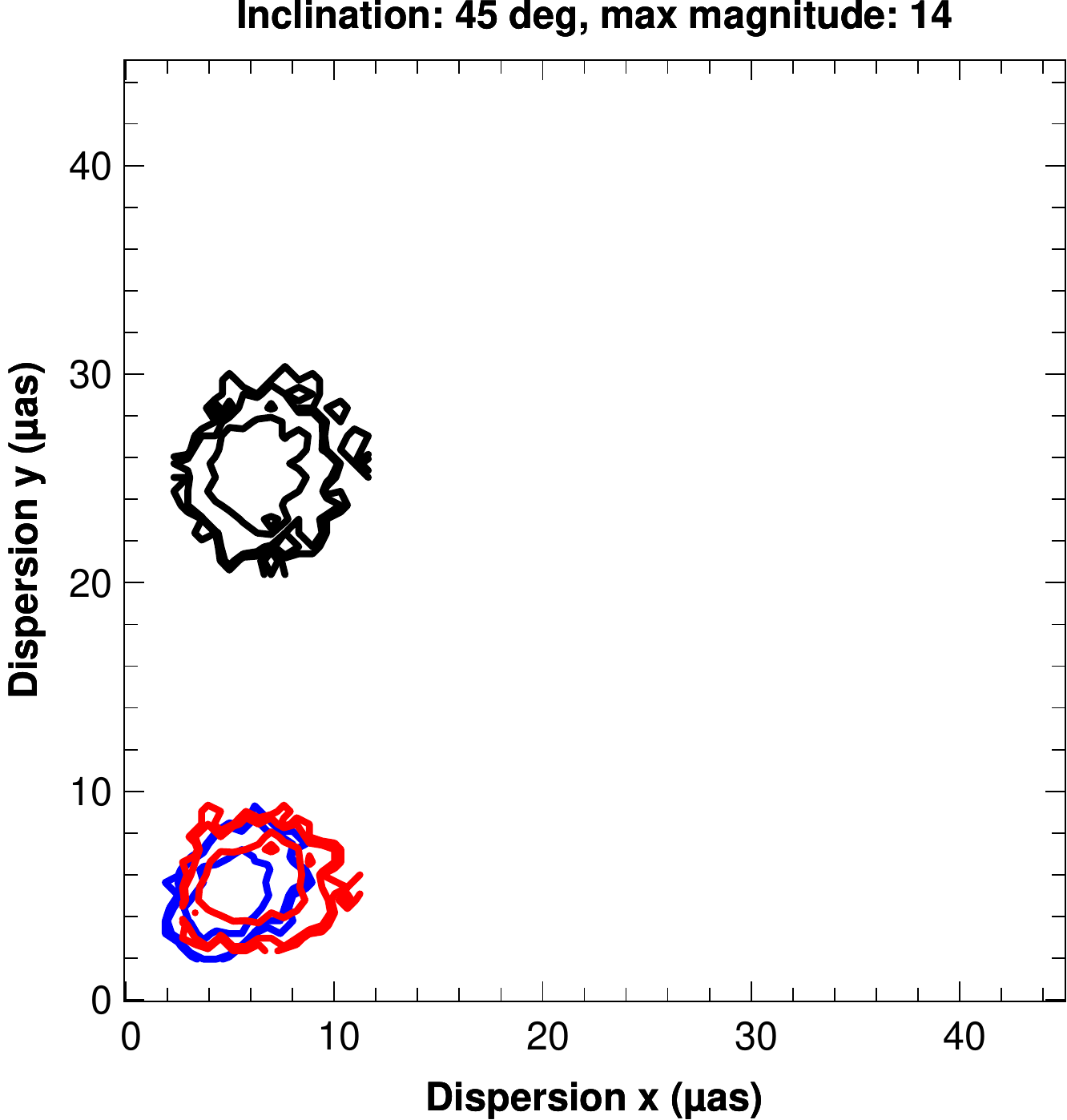}
	\includegraphics[width=5cm,height=5cm]{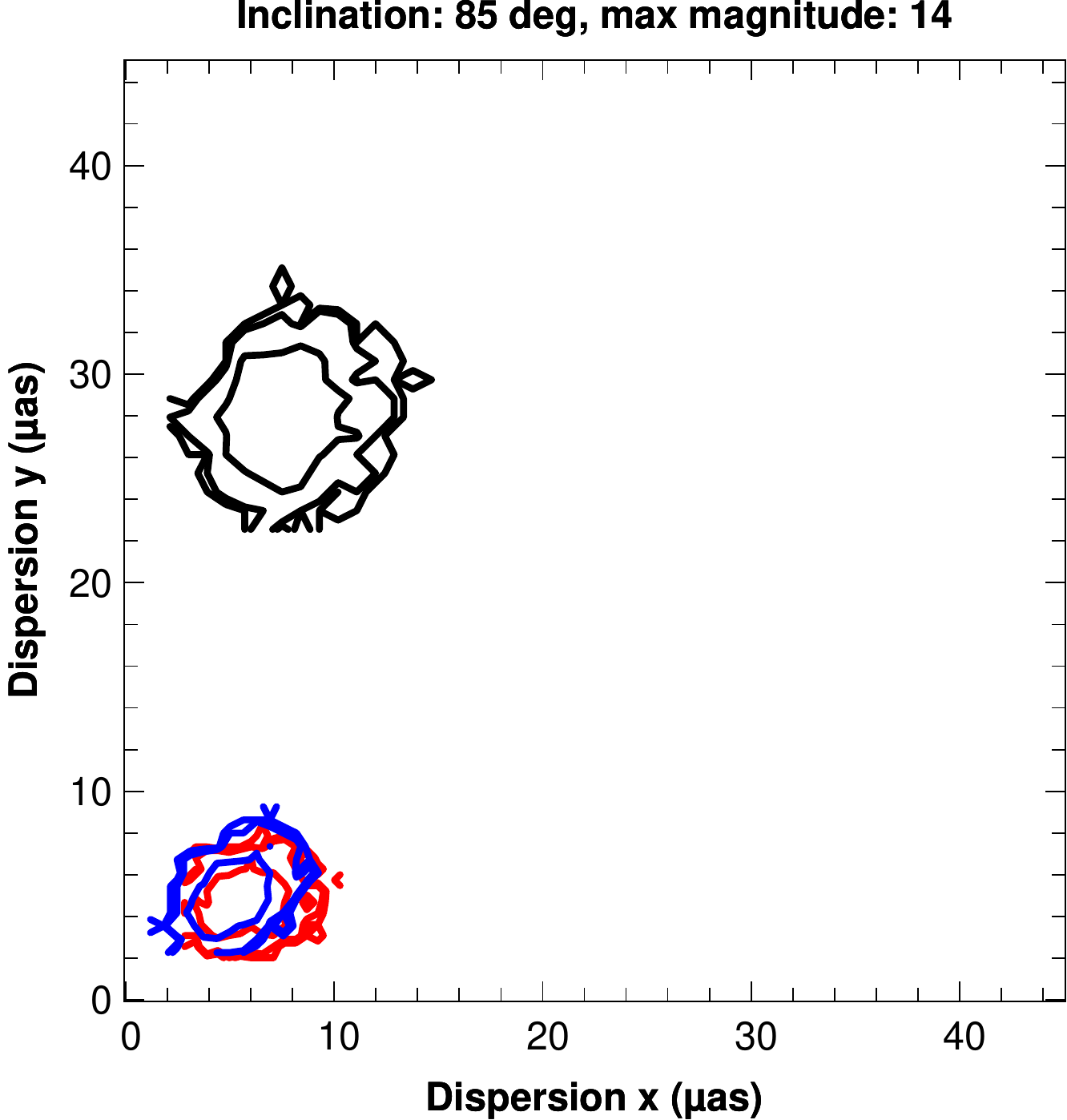}
	\caption{Histograms of directional dispersions $(\sigma_x,\sigma_y)$ (defined in Eq.~\ref{eq:disp}) of the \textit{GRAVITY} astrometric data
		retrieved in the framework of the RWI (blue), RN (red) and blob (black) models. The inclination
		is $5^{\circ}$ (left), $45^{\circ}$ (centre) or $85^{\circ}$ (right). The brightest
		magnitude of the flare is always $m_{\mathrm{K}}=14$. 
		The contours encompass respectively around
		$68\%$, $95\%$ and $99\%$ of the simulated $1000$ nights.The blob model is easily distinguished from
		the two other models at medium and high inclination.}
	\label{fig:histo1d14}
\end{figure*}

\begin{figure*}
\centering
	\includegraphics[width=5cm,height=5cm]{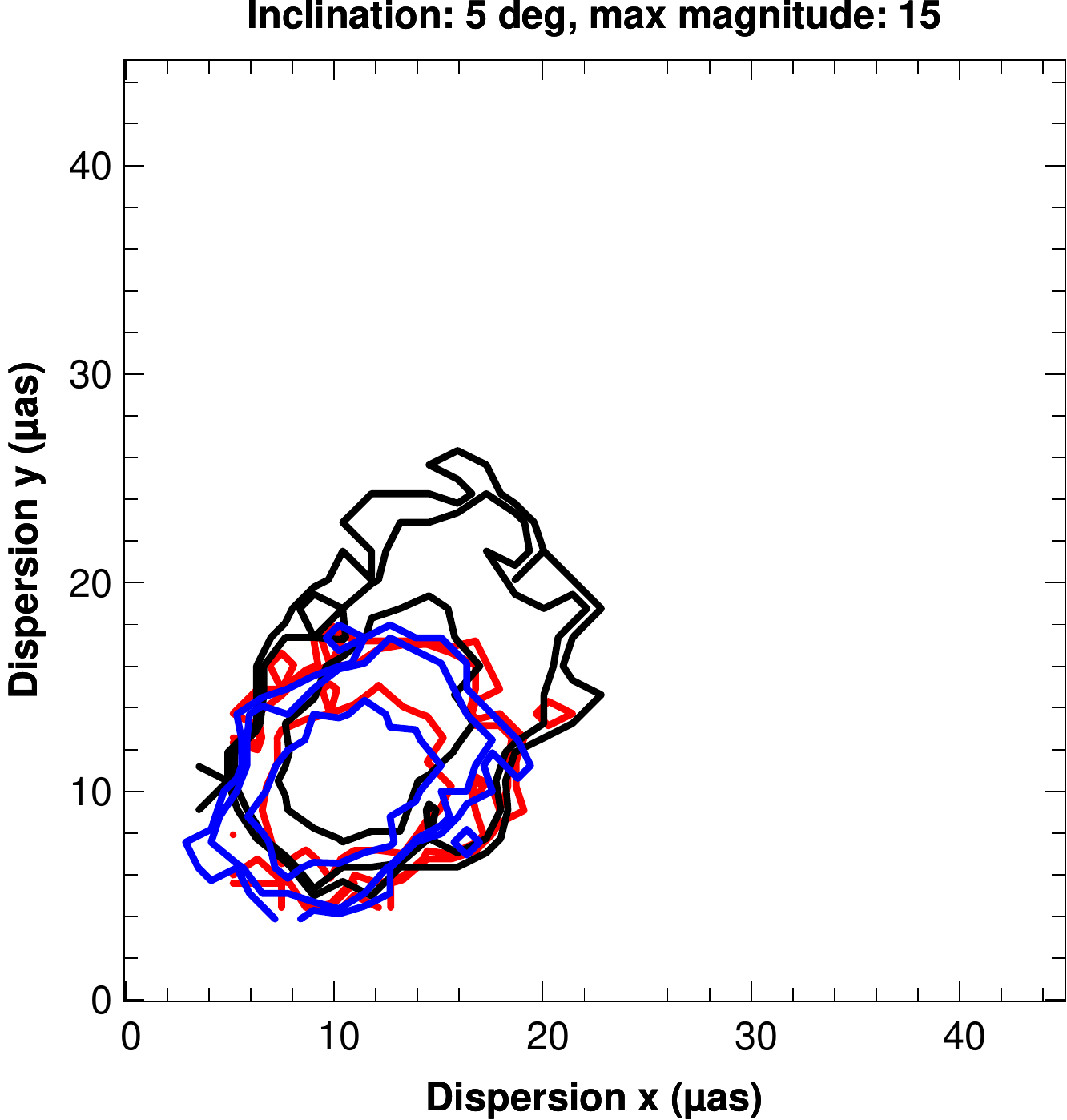}
	\includegraphics[width=5cm,height=5cm]{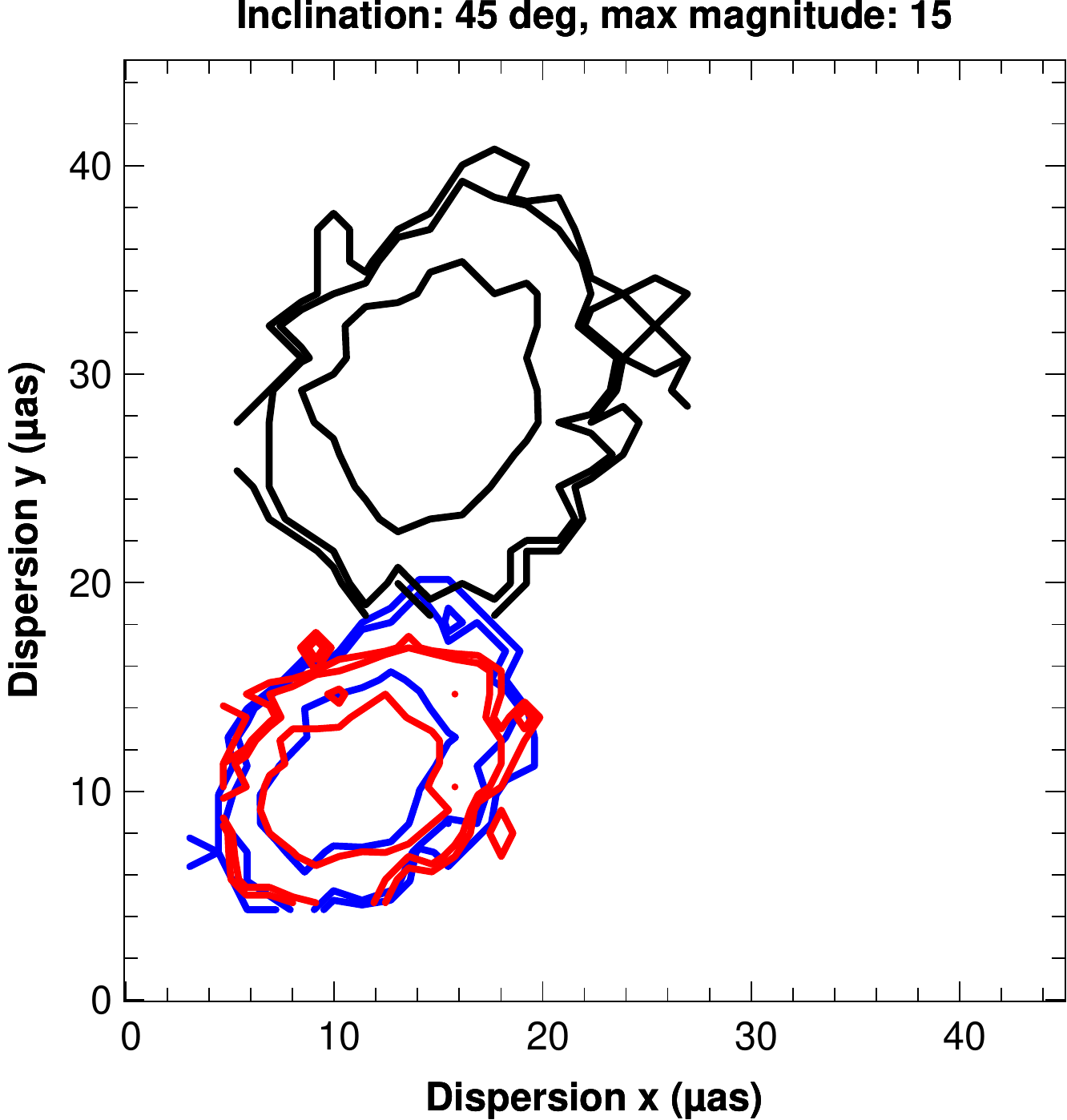}
	\includegraphics[width=5cm,height=5cm]{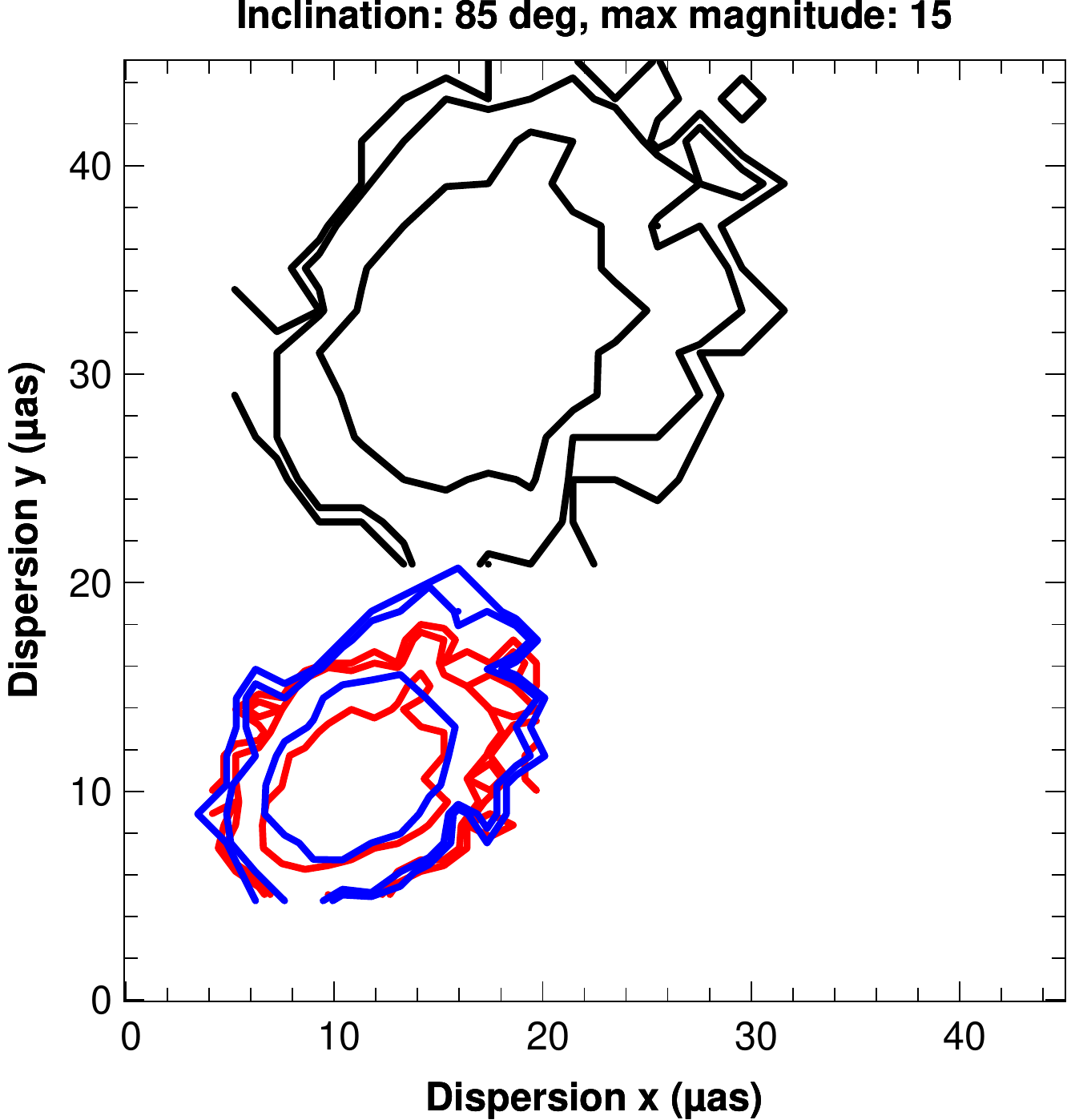}
	\caption{Same as Fig.~\ref{fig:histo1d14} with $m_{\mathrm{K}}=15$.
	The blob model is still easily distinguished from
		the two other models at medium and high inclination.}
	\label{fig:histo1d15}
\end{figure*}

Figs.~\ref{fig:histo1d14} and~\ref{fig:histo1d15} show the histograms of 
directional dispersion of the RWI, RN and blob models
when the inclination and brightest magnitude of the flare are varied in $i\in[5^{\circ};45^{\circ};85^{\circ}]$
and $m_{\mathrm{K}}\in[14;15]$.

The RWI and RN are not distinguishable, even if many bright $(m_{\mathrm{K}}=14)$ 
and long ($2$~h) flares were observed,
as their respective histograms are nearly superimposed. This holds whatever the inclination.
However, the blob model histograms are very
clearly separated from the two other models at medium and high inclination, both
for a brightest flare magnitude of $m_{\mathrm{K}}=14$ and $m_{\mathrm{K}}=15$.
At small inclination ($i\approx5^{\circ}$), all models are indistinguishable.



\begin{figure*}
\centering
	\includegraphics[width=5cm,height=5cm]{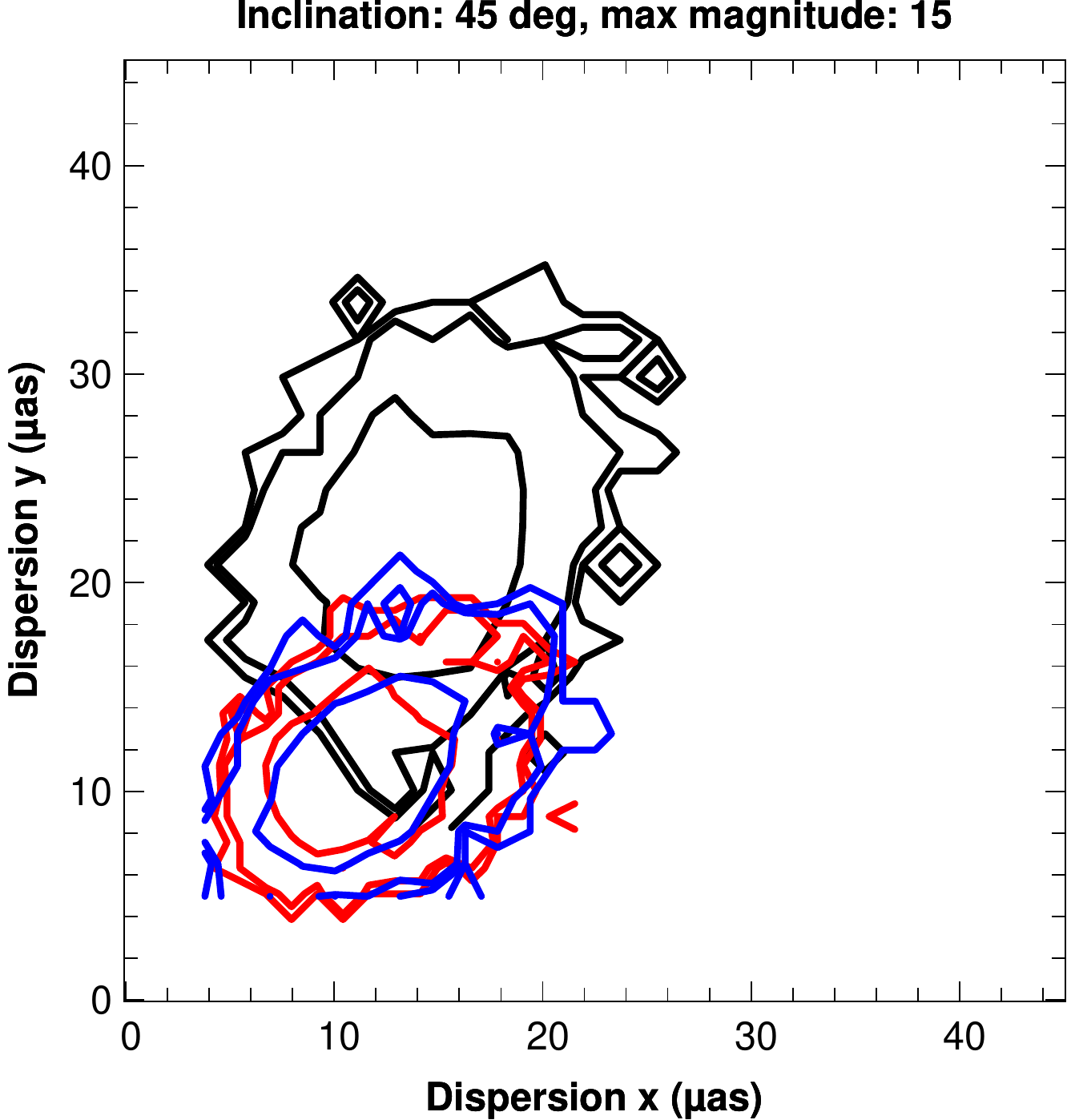}
	\includegraphics[width=5cm,height=5cm]{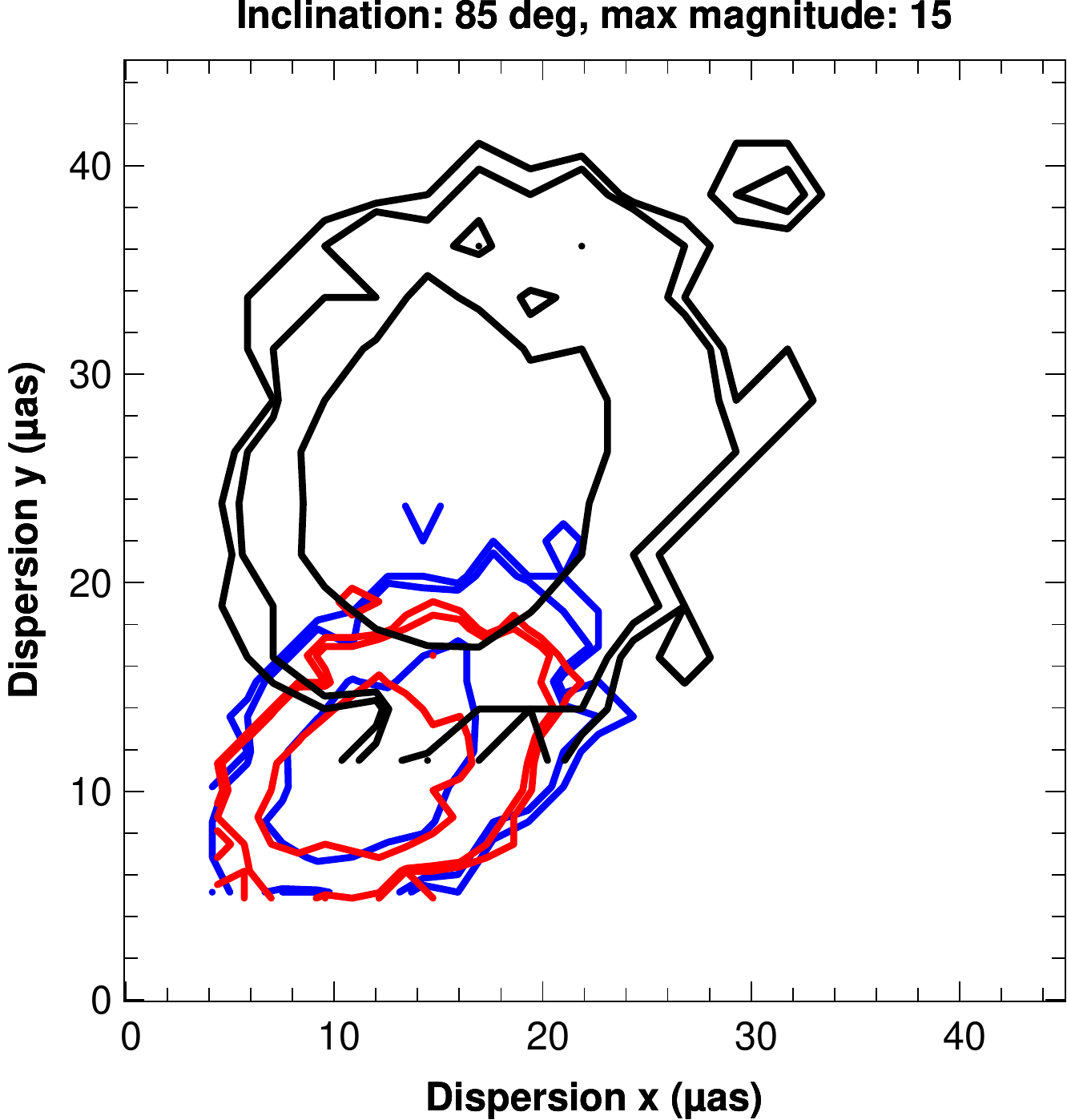}
	\caption{Histograms of the directional dispersion for a \textit{shorter flare that lasts approximately $1$~h~$30$}
	(as compared to $2$~h for Fig.~\ref{fig:histo1d14}~-~\ref{fig:histo1d15}),
		with a brightest magnitude of
		$m_{\mathrm{K}}=15$ and an inclination 
		of $45^{\circ}$ (left) or $85^{\circ}$ (right). The distinction of the blob model is still possible.}
	\label{fig:histo2dmedium}
\end{figure*}

\begin{figure*}
\centering
	\includegraphics[width=5cm,height=5cm]{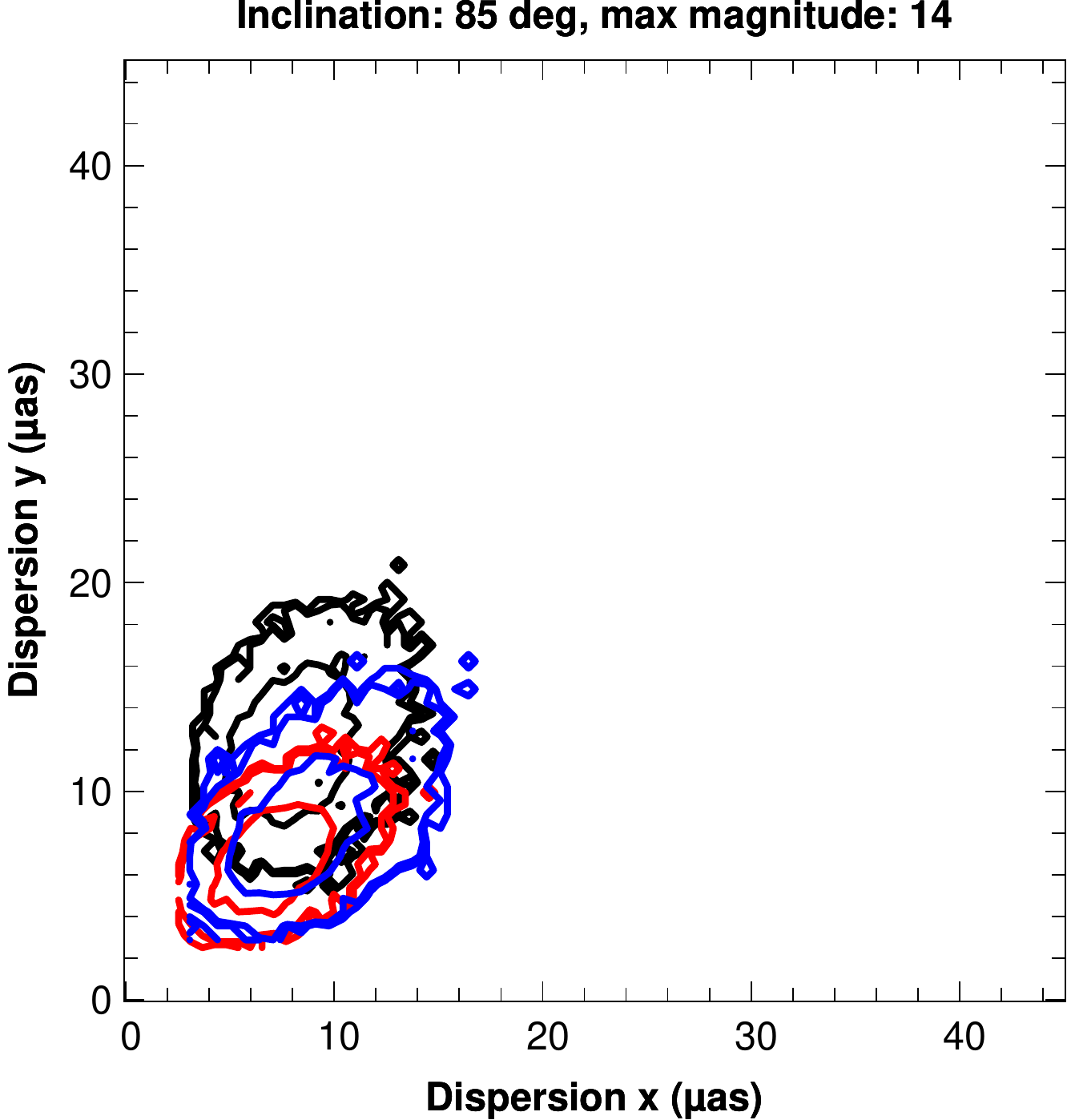}
	\caption{Histograms of the directional dispersion for an \textit{even shorter flare that lasts approximately $1$~h}
	(as compared to $2$~h for Figs.~\ref{fig:histo1d14}~-~\ref{fig:histo1d15}),
		with a brightest magnitude of
		$m_{\mathrm{K}}=14$ and an inclination 
		of $85^{\circ}$. The distinction of the blob model is still marginally possible.}
	\label{fig:histo2dshort}
\end{figure*}

It is interesting to determine whether the distinction between the blob
models and its alternatives is strongly dependent on the flare duration.
Fig.~\ref{fig:histo2dmedium} shows the directional dispersion obtained
when the flare duration is reduced from $2$~h to $1$~h~$30$, for a brightest
magnitude of $m_{\mathrm{K}}=15$. It shows that
the blob model can still be easily distinguished. In order to model a shorter
flare in the framework of the blob model, the ejection velocity has been
reduced from $0.4~c$ to $0.36~c$, leading to a smaller maximum altitude
of the blob ($60~\mu$as as compared to $100~\mu$as, at $45^{\circ}$ of inclination).
When reducing the flare duration (and hence the blob ejection velocity) to approximately
$1$~h, it becomes very difficult to distinguish the models. In this case, the blob's motion has
a vertical extension of around $20~\mu$as, which gets very close to the typical
extension of the centroid motion for RWI and RN models. Fig.~\ref{fig:histo2dshort} shows
that even for a brightest magnitude of $m_{\mathrm{K}}=14$ and an inclination of $85^{\circ}$,
the directional dispersion of the three models overlap. However, the distinction is still
marginally possible in this case, but many events would be necessary to disentangle
clearly the models.

\section{Conclusion}
\label{sec:conclu}

We have analysed the ability of the near-future \textit{GRAVITY} instrument
to distinguish different flare models, depending only on their astrometric
signatures. We show that the ejection of a blob can be distinguished from
alternative models provided the ejection direction is not face-on for the observer.
This result holds for a reasonably long flare ($\gtrsim 1$~h~$30$) with a brightest
magnitude in {\it K}-band of $14 \lesssim m_{\mathrm{K}} \lesssim 15$.

%


{Our models are very simple for the time being. In particular, we use a geometrically thin
disk approximation with very simple emission laws and a pseudo-Newtonian potential
for all MHD simulations. However, this simple framework allows to generate images of
Sgr~A* accretion flow that looks rather similar to more complex simulations. 
}

{The main result of our simulations is that \textit{GRAVITY} will be able to distinguish an ejection
model as compared to "disk-glued" models. This result is rather natural, as most
models of disk-like (be it geometrically thin or thick) accretion structures will give rise to typical
crescent-shape images~\citep{kamruddin13}, and will lead to rather similar
centroid tracks. We believe that this result is robust and will not be changed
by developing more sophisticated simulations.}


Our result is important in the sense that it demonstrates the ability
of \textit{GRAVITY} to make an observable difference between 
categories of flare models, which has not been possible so far with
current instrumentation. 

Future work will be dedicated to determining to what extent can
the astrometric data of \textit{GRAVITY} help distinguish models of the
two first astrometric signature classes, as defined in Section~\ref{sec:intro} (typically,
the red-noise and RWI models). This will demand resorting not only
on astrometry but on other signatures as well, such as photometry
and polarimetry. Let us repeat here that our current results,
allowing to distinguish an ejected blob, only use astrometric data,
thus only a part of the total flare data.

Future work will also be dedicated to modelling various
flare models using GRMHD simulations
in order to be able to study the impact of the spin parameter on the observables.
Such a study of Sgr~A* variability in GRMHD simulations is currently developed
by various groups~{\citep[see e.g.][]{dexter09,dolence12,henisey12,dexter13,shcherbakov13}}.

\section*{Acknowledgments}
FHV and PV acknowledge financial support from the French Programme National des Hautes \'Energies (PNHE).
PV and FC acknowledge financial
support from the UnivEarthS Labex program at Sorbonne Paris Cit\'e
(ANR-10-LABX-0023 and ANR-11-IDEX-0005-02).
Computing was partly done using the Division Informatique de
l'Observatoire (DIO) HPC facilities from Observatoire de Paris
(\url{http://dio.obspm.fr/Calcul/}).
Calculations were partially performed at the FACe (Fran\c cois Arago
Centre) in Paris.

\bibliographystyle{mn2e}
\bibliography{biblio}

\appendix
\label{appendix}

\section{Detection noise for \textit{GRAVITY} observation}
\label{noise}

For completeness we discuss here the sources of detection noise that will affect \textit{GRAVITY} observations.
This discussion comes from~\citet{vincent11}.

\textit{GRAVITY} will use the four Very Large Telescope Unit Telescopes (UT) to observe Sgr~A*. Each baseline will combine the beams of intensities $I_{p}$ and $I_{q}$ delivered by two telescopes labelled $p$ and $q$. Let the intrinsic visibility modulus and phase of the observed object be $V_{\mathrm{obj}}$ and $\phi_{\mathrm{obj}}$. Let $\delta \phi$ be the phase shift between the two channels from telescope $p$ and $q$. The short-exposure noiseless combined intensity is:

\begin{equation}
	\label{Icomb}
	I(\delta \phi) =  I_{p}+I_{q}+2\sqrt{I_{p} I_{q}}V_{\mathrm{obj}} \mathrm{cos}(\delta \phi + \phi_{\mathrm{obj}} + \phi_{\mathrm{piston,short}})
\end{equation}
where $\phi_{\mathrm{piston,short}}$ is the short-exposure piston phase.

It is sufficient to get four samples of this function in order to retrieve the visibility modulus and the phase of the object. Indeed, if the four following quantities are computed:

\begin{eqnarray}
	\label{ABCD}
A  &=& I(0), \\ \nonumber
B  &=& I(\frac{\pi}{2}), \\ \nonumber
C  &=& I(\pi), \\ \nonumber
D  &=& I(\frac{3\,\pi}{2}), \nonumber
\end{eqnarray}

then it is easy to show that the complex visibility of the object is given by:

\begin{equation}
	\label{visABCD}
	V_{\mathrm{obj}} e^{i\,\phi_{\mathrm{obj}}} = 2\frac{(A-C)-i\,(B-D)}{A+B+C+D}.
\end{equation}

In order to be able to simulate realistic quantities, we now take into account the various sources of noises affecting the different quantities that have been introduced.

Let $N_{\rm{ph}}$ be the number of photons arriving from each of the $N_{\rm{tel}}$ telescopes of the interferometer. Each set of $N_{\rm{ph}}$ photons will be dispatched to the $N_{\rm{tel}}-1$ other telescopes. Then $\frac{2N_{\rm{ph}}}{N_{\rm{tel}}-1}$ photons will be present in the baseline between two telescopes. Let  $T$ be the transmission of the instrument multiplied by the quantum yield, estimated to be 0.009. The mean number of photo-electrons $\langle m \rangle$ per sample per baseline is:

\begin{equation}
	\label{photon}
	\langle m \rangle=\frac{2N_{\rm{ph}}T}{4(N_{\rm{tel}}-1)}.
\end{equation}

Hence the shot noise: $\sigma_{\mathrm{shot}}=\sqrt{\langle m \rangle}$. It is assumed that the number of read out noise electrons is equal to $\sigma_{\mathrm{RON}}^{2}=36$. Given the rate of dark current electrons $N_{\mathrm{dark}} = 100\; \rm{s}^{-1}$, the variance of the detection noise per integration time $\tau$ is:

\begin{equation}
	\label{sigmadetec}
	\sigma^{2}_{\mathrm{detec}}=\sigma_{\mathrm{shot}}^{2}+\sigma_{\mathrm{RON}}^{2}+N_{\mathrm{dark}}\,\tau.
\end{equation}

Taking this noise into account, and assuming a gaussian distribution for the various noise contributions, it is possible to compute a realization of the detection noise $n_{A},n_{B},n_{C},n_{D}$ corrupting the four A,B,C,D signals, hence the noise on the object complex visibility.

\label{lastpage}

\end{document}